\documentstyle[11pt]{article}
\newcommand{\be}{\begin{equation}}
\newcommand{\ee}{\end{equation}}
\newcommand{\bea}{\begin{array}}
\newcommand{\eea}{\end{array}}

\title{$(X,\psi)$ DUALITY and ENHANCED dKdV\\
ON A RIEMANN SURFACE}
\author{Robert Carroll\\
Mathematics Department\\
University of Illinois\\
Urbana, IL 61801\\email:  rcarroll@symcom.math.uiuc.edu}

\date{May, 1997}

\begin{document}
\bibliographystyle{plain}
\maketitle

\begin{abstract} 
Given a dKdV potential $V$, arising from a finite zone KdV situation on a
Riemann surface $\Sigma$, one can create an enhanced dispersionless
context $dKdV_{\epsilon}$ with an expanded $V$ (retaining powers of 
$\epsilon$) in which various formulas in the $(X,\psi)$ duality theory
of Faraggi-Matone in \cite{fa} have representations, and a natural
symplectic form $(dX/\epsilon)\wedge dQ\,\,(P=iQ)$ in the Hamilton-Jacobi
theory for dKdV has a representation in terms of the prepotential 
${\cal F}$ of \cite{fa}.  The theory establishes relations between
an expanded ${\cal F}_{\epsilon}=\tilde{{\cal F}}$
and the free energy $F_{\epsilon}$ of $dKdV_{\epsilon}$ which lead to
formulas relating the duality variables $a_i,\,\,
a^D_i$ of Seiberg-Witten type on $\Sigma$ to $\tilde{Q}=\Im\tilde{P}=
-(1/2\Re\tilde{{\cal F}})$.  Formulas at various stages of truncation
in $\epsilon$ are also indicated and they usually involve constraints
on $Q$ for example.
\end{abstract}


\section{INTRODUCTION}
\renewcommand{\theequation}{1.\arabic{equation}}\setcounter{equation}{0}

The dispersionless theory of KP (Kadomtsev-Petviashvili) and Toda
hierarchies, along with connections to Whitham equations, topological
field theory (TFT), and Seiberg-Witten (SW) theory, has been extensively
developed in recent years (see. e.g. \cite{aa,bb,ca,cb,cc,cd,ce,cf,
cg,ch,ci,cj,ck,cl,da,dc,dd,fc,fb,ga,ia,ke,ka,kb,kc,kf,na,ta} and 
further references there).  In \cite{cb,ck}, developing the theme
of \cite{ca}, we were led to an extension of dispersionless theory
(denoted by $dKP_{\epsilon}$ or $dKdV_{\epsilon}$ there) in order
to connect the $(X,\psi)$ duality theory of \cite{fa} to WKB type
formulas.  We will review these connections here and give a more systematic
development of the enhanced dispersionless theory in relation
to $(X,\psi)$ duality.
For convenience we restrict ourselves fo KP and KdV (Korteweg-deVries)
situations since they will naturally arise in connections to \cite{fa}
and we will only give a full exposition for KdV after showing that this
is the appropriate theory.  The equations for extended (or enhanced)
dKP here (and for other hierarchies) are then easily written down
and we omit details.  This paper is essentially a refinement of 
\cite{ck}, with extraction of material from \cite{ca,cb}, plus
a few new ideas.

\section{BACKGROUND FOR DISPERSIONLESS THEORY}
\renewcommand{\theequation}{2.\arabic{equation}}\setcounter{equation}{0}

\subsection{Classical framework for KP}

We give next as in \cite{cb}
a brief sketch of some ideas regarding dispersionless KP
(dKP) following mainly \cite{ch,ci,cj,ke,ta} to which we refer for 
philosophy.  We will make various notational adjustments as we go along 
and subsequently will modify some of the theory.  One
can think of fast and slow variables with $\epsilon x=X$ and $\epsilon t_n=
T_n$ so that $\partial_n\to\epsilon\partial/\partial T_n$ and $u(x,t_n)
\to\tilde{u}(X,T_n)$ to obtain from the KP equation $(1/4)u_{xxx}+3uu_x
+(3/4)\partial^{-1}\partial^2_2u=0$ the equation $\partial_T\tilde{u}=3
\tilde{u}\partial_X\tilde{u}+(3/4)\partial^{-1}(\partial^2\tilde{u}/
\partial T_2^2)$ when $\epsilon\to 0$ ($\partial^{-1}\to(1/\epsilon)
\partial^{-1}$).  In terms of hierarchies the theory can be built around the
pair $(L,M)$ in the spirit of \cite{ci,cm,ta}.  Thus writing $(t_n)$ for
$(x,t_n)$ (i.e. $x\sim t_1$ here) consider
\be
L_{\epsilon}=\epsilon\partial+\sum_1^{\infty} u_{n+1}(\epsilon,T)
(\epsilon\partial)^{-n};\,\,M_{\epsilon}=\sum_1^{\infty}nT_nL^{n-1}_{\epsilon}
+\sum_1^{\infty}v_{n+1}(\epsilon,T)L_{\epsilon}^{-n-1}
\label{YA}
\ee
Here $L$ is the Lax operator $L=\partial+\sum_1^{\infty}u_{n+1}\partial^{-n}$
and $M$ is the Orlov-Schulman operator defined via $\psi_{\lambda}=M\psi$.
Now one assumes $u_n(\epsilon,T)=U_n(T)+O(\epsilon)$, etc. and 
sets (recall $L\psi=\lambda\psi$)
$$
\psi=\left[1+O\left(\frac{1}{\lambda}\right)\right]exp
\left(\sum_1^{\infty}\frac{T_n}
{\epsilon}\lambda^n\right)=exp\left(\frac{1}
{\epsilon}S(T,\lambda)+O(1)\right);$$
\be
\tau=exp\left(\frac{1}{\epsilon^2}F(T)+O\left(\frac{1}{\epsilon}\right)
\right)
\label{YB}
\ee
We recall that $\partial_nL=[B_n,L],\,\,B_n=L^n_{+},\,\,\partial_nM
=[B_n,M],\,\,[L,M]=1,\,\,L\psi=\lambda\psi,\,\,\partial_{\lambda}\psi
=M\psi,$ and $\psi=\tau(T-(1/n\lambda^n))exp[\sum_1^{\infty}T_n\lambda^n]/
\tau(T)$.  Putting in the $\epsilon$ and using $\partial_n$ for
$\partial/\partial T_n$ now, with $P=S_X$, one obtains
\be
\lambda=P+\sum_1^{\infty}U_{n+1}P^{-n};\,\,
P=\lambda-\sum_1^{\infty}P_i\lambda^{-1};
\label{YC}
\ee
$${\cal
M}=\sum_1^{\infty}nT_n\lambda^{n-1}+\sum_1^{\infty}V_{n+1}\lambda^{-n-1};
\,\,\partial_nS={\cal B}_n(P)\Rightarrow \partial_nP=\hat{\partial}
{\cal B}_n(P)$$
where $\hat{\partial}\sim \partial_X+(\partial P/\partial X)\partial_P$
and $M\to {\cal M}$
(note that one assumes also $v_{i+1}(\epsilon,T)=V_{i+1}(T)+O(\epsilon)$). 
Further
for $B_n=\sum_0^nb_{nm}\partial^m$ one has ${\cal B}_n=\sum_0^nb_{nm}
P^m$ (note also $B_n=L^n+\sum_1^{\infty}\sigma_j^nL^{-j}$).  
We list a few additional formulas which are 
easily obtained (cf. \cite{ci}); thus, writing $\{A,B\}=\partial_PA\partial A
-\partial A\partial_PB$ one has
\be
\partial_n\lambda=\{{\cal B}_n,\lambda\};\,\,\partial_n{\cal M}
=\{{\cal B}_n,{\cal M}\};\,\,\{\lambda,{\cal M}\}=1
\label{YD}
\ee
Now we can write $S=\sum_1^{\infty}T_n\lambda^n+\sum_1^{\infty}S_{j+1}
\lambda^{-j}$ with $S_{n+1}=-(\partial_nF/n),\,\,
\partial_mS_{n+1}=(F_{mn}/n),
\,\,V_{n+1}=-nS_{n+1}$,
and $\partial_{\lambda}S={\cal M}$.
Further 
\be
{\cal B}_n=\lambda^n+\sum_1^{\infty}\partial_nS_{j+1}\lambda^{-j};\,\,
\partial S_{n+1}\sim -P_n\sim -\frac{\partial V_{n+1}}{n}\sim
-\frac{\partial\partial_n F}{n}
\label{YE}
\ee
\indent
We sketch next a few formulas from \cite{ke} (cf. also \cite{ci}).  First
it will be important to rescale the $T_n$ variables and write 
$t'=nt_n,\,\,T_n'=nT_n,\,\,
\partial_n=n\partial'_n=n(\partial/\partial T'_n)$.  Then
\be
\partial'_nS=\frac{\lambda^n_{+}}{n};\,\,\partial'_n\lambda=\{{\cal Q}_n,
\lambda\}\,\,({\cal Q}_n=\frac{{\cal B}_n}{n});
\label{YF}
\ee
$$\partial'_nP=\hat{\partial}{\cal Q}_n=\partial{\cal Q}_n+\partial_P
{\cal Q}_n\partial P;\,\,\partial'_n{\cal Q}_m-\partial'_m{\cal Q}_n=
\{{\cal Q}_n,{\cal Q}_m\}$$
Now think of $(P,X,T'_n),\,\,n\geq 2,$ as basic Hamiltonian variables
with $P=P(X,T'_n)$.  Then $-{\cal Q}_n(P,X,T'_n)$ will serve as a
Hamiltonian via
\be
\dot{P}'_n=\frac{dP'}{dT'_n}=\partial{\cal Q}_n;\,\,\dot{X}'_n=\frac
{dX}{dT'_n}=-\partial_P{\cal Q}_n
\label{YG}
\ee
(recall the classical theory for variables $(q,p)$ involves $\dot{q}=
\partial H/\partial p$ and $\dot{p}=-\partial H/\partial q$).  The function
$S(\lambda,X,T_n)$ plays the role of part of a generating function 
$\hat{S}$ for the Hamilton-Jacobi theory with action angle variables 
$(\lambda,-\xi)$ where
\be
PdX+{\cal Q}_ndT'_n=-\xi d\lambda-K_ndT'_n+d\hat{S};\,\,K_n=-R_n=-
\frac{\lambda^n}{n};
\label{YH}
\ee
$$\frac{d\lambda}{dT'_n}=\dot{\lambda}'_n=\partial_{\xi}R_n=0;\,\,\frac
{d\xi}{dT'_n}=\dot{\xi}'_n=-\partial_{\lambda}R_n=-\lambda^{n-1}$$
(note that $\dot{\lambda}'_n=0\sim\partial'_n\lambda=\{{\cal Q}_n,
\lambda\}$).  To see how all this fits together we write
\be
\frac{dP}{dT'_n}=\partial'_nP+\frac{\partial P}{\partial X}\frac{dX}
{dT'_n}=\hat{\partial}{\cal Q}_n+\frac{\partial P}{\partial X}\dot
{X_n}'=\partial{\cal Q}_n+\partial P\partial_P{\cal Q}_n+\partial P
\dot{X}'_n
\label{YI}
\ee
This is compatible with (\ref{YG}) and Hamiltonians $-{\cal Q}_n$.  Furthermore
one wants
\be
\hat{S}_{\lambda}=\xi;\,\,\hat{S}_X=P;\,\,\partial'_n\hat{S}=
{\cal Q}_n-R_n
\label{YJ}
\ee
and from (\ref{YH}) one has
\be
PdX+{\cal Q}_ndT'_n=-\xi d\lambda+R_ndT'_n+\hat{S}_XdX+\hat{S}_
{\lambda}d\lambda+\partial'_n\hat{S}dT'_n
\label{YK}
\ee
which checks.  We note that $\partial'_nS={\cal Q}_n={\cal B}_n/n$ and
$S_X=P$ by constructions and definitions.  Consider $\hat{S}=S-\sum_2^{\infty}
\lambda^nT'_n/n$.  Then $\hat{S}_X=S_X=P$ and $\hat{S}_n'=S_n'-R_n=
{\cal Q}_n-R_n$ as desired with $\xi=\hat{S}_{\lambda}=S_{\lambda}-
\sum_2^{\infty}T'_n\lambda^{n-1}$.  It follows that
$\xi\sim{\cal M}-\sum_2^{\infty}T'_n\lambda^{n-1}=X+\sum_1^{\infty}V_{i+1}
\lambda^{-i-1}$.  If $W$ is the gauge operator such that $L=W\partial W^{-1}$
one sees easily that
\be
M\psi
=W\left(\sum_1^{\infty}kx_k\partial^{k-1}\right)W^{-1}\psi=\left(G+
\sum_2^{\infty}kx_k\lambda^{k-1}\right)\psi
\label{YL}
\ee
from which follows that $G=WxW^{-1}\to\xi$.  This shows that $G$ is
a very fundamental object and this is encountered in various places
in the general theory (cf. \cite{ci,cm}).

\subsection{Dispersonless theory for KdV}

Following \cite{ci,co,cp} we write
\be
L^2=L^2_{+}=\partial^2+q=\partial^2-u\,\,(q=-u=2u_2);\,\,q_t-6qq_x-q_{xxx}=0;
\label{YM}
\ee
$$B=4\partial^3+6q\partial+3q_x;\,\,L^2_t=[B,L^2];\,\,q=-v^2-v_x\sim
v_t+6v^2v_x+v_{xxx}=0$$
($v$ satisfies the mKdV equation).  Canonical formulas would involve
$B\sim B_3=L^3_{+}$ as indicated below but we retain the $B$ momentarily
for comparison to other sources. 
KdV is Galilean invariant ($x'=x-
6\lambda t,\,\,t'=t,\,\,u'=u+\lambda$) and consequently one can consider
$L+\partial^2+q-\lambda=(\partial+v)(\partial-v,\,\,q-\lambda=-v_x-v^2,\,\,
v=\psi_x/\psi,$ and $-\psi_{xx}/\psi=q-\lambda$ or $\psi_{xx}+q\psi=\lambda\psi$
(with $u'=u+\lambda\sim q'=q-\lambda$).  The $v$ equation in (\ref{YM}) becomes
then $v_t=\partial(-6\lambda v+2v^3-v_{xx})$ and for $\lambda=-k^2$ one
expands for $\Im k>0,\,\,|k|\to\infty$ to get
$(\bullet)\,\,
v\sim ik+\sum_1^{\infty}(v_n/(ik)^n)$.  The $v_n$ are conserved
densities and with $2-\lambda=-v_x-v^2$ one obtains
\be
p=-2v_1;\,\,2v_{n+1}=-\sum_1^{n-1}v_{n-m}v_m-v'_n;\,\,2v_2=-v'_1
\label{YN}
\ee
Next for $\psi''-u\psi=-k^2\psi$
write $\psi_{\pm}\sim exp(\pm ikx)$ as $x\to\pm\infty$.  Recall also the 
transmission and reflection coefficient formulas (cf. \cite{ce})
$T(k)\psi_{-}=R(k)\psi_{+}+\psi_{+}(-k,x)$ and $T\psi_{+}=R_L\psi_{-}+
\psi_{-}(-k,x)$.  Writing e.g. $\psi_{+}=exp(ikx+\phi(k,x))$ with
$\phi(k,\infty)=0$ one has $\phi''+2ik\phi'+(\phi')^2=u$.  Then
$\psi'_{+}/\psi_{+}=ik+\phi'=v$ with $q-\lambda=-v_x-v^2$.  Take then
\be
\phi'=\sum_1^{\infty}\frac{\phi_n}{(2ik)^n};\,\,v\sim ik+\phi'=ik+
\sum\frac{v_n}{(ik)^n}\Rightarrow\phi_n=2^nv_n
\label{YO}
\ee
Furthermore one knows 
\be
log T=-\sum_0^{\infty}\frac{c_{2n+1}}{k^{2n+1}};\,\,c_{2n+1}=\frac{1}{2\pi i}
\int_{-\infty}^{\infty}k^{2n}log(1-|R|^2)dk
\label{YP}
\ee
(assuming for convenience that there are no bound states).  Now for
$c_{22}=R_L/T$ and $c_{21}=1/T$ one has as $k\to -\infty\,\,(\Im k>0)$ the
behavior $\psi_{+}exp(-ikx)\to c_{22}exp(-2ikx)+c_{21}\to c_{21}$.  Hence
$exp(\phi)\to c_{21}$ as $x\to -\infty$ or $\phi(k,-\infty)=-log T$ which
implies
\be
\int_{-\infty}^{\infty}\phi'dx=log T=\sum_1^{\infty}\int_{-\infty}^{\infty}
\frac{\phi_ndx}{(2ik)^n}
\label{YQ}
\ee
Hence $\int\phi_{2m}dx=0$ and $c_{2m+1}=-\int\phi_{2m+1}dx/(2i)^{2m+1}$.
The $c_{2n+1}$ are related to Hamiltonians $H_{2n+1}=\alpha_nc_{2n+1}$
as in \cite{cm,cp} and thus the conserved densities $v_n\sim \phi_n$ give
rise to Hamiltonians $H_n$ (n odd).  There are action angle variables
$P=klog|T|$ and $Q=\gamma arg(R_L/T)$ with Poisson structure $\{F,G\}\sim
\int(\delta F/\delta u)\partial (\delta G/\delta u)dx$ (we omit the
second Poisson structure here).  
\\[3mm]\indent
Now look at the dispersionless theory based on $k$ where $\lambda^2
\sim(ik)^2=-k^2$.  One obtains for $P=S_X,\,\,P^2+q=-k^2$, and we write
${\cal P}=(1/2)P^2+p=(1/2)(ik)^2$ with $q\sim 2p\sim 2u_2$.  One has
$\partial k/\partial T_{2n}=\{(ik)^{2n},k\}=0$ and from $ik=P(1+qP^{-2})^
{1/2}$ we obtain
\be
ik=P\left(1+\sum_1^{\infty}{\frac{1}{2}\choose m}q^mP^{-2m}\right)
\label{YR}
\ee
(cf. (\ref{YC}) with $u_2=q/2$).  The flow equations become then
\be
\partial'_{2n+1}P=\hat{\partial}{\cal Q}_{2n+1};\,\,\partial'_{2n+1}(ik)
=\{{\cal Q}_{2n+1},ik\}
\label{YS}
\ee
Note here some rescaling is needed since we want $(\partial^2+q)^{3/2}_{+}=
\partial^3+(3/2)q\partial+(3/4)q_x=B_3$ instead of our previous $B_3\sim
4\partial^3 +6q\partial+3q_x$.  Thus we want ${\cal Q}_3=(1/3)P^3+(1/2)qP$
to fit the notation above.  The Gelfand-Dickey resolvant 
coefficients are defined via $R_s(u)=(1/2)Res(\partial^2-u)^{s-(1/2)}$
and in the dispersionless picture 
$R_s(u)\to (1/2)r_{s-1}(-u/2)$ (cf. \cite{ci}) where
$$
r_n=Res(-k^2)^{n+(1/2)}=\left(
\begin{array}{c}
n+(1/2)\\
n+1
\end{array}\right)q^{n+1}=
\frac{(n+1/2)\cdots(1/2}{(n+1)!}q^{n+1};$$
\be
\,\,2\partial_qr_n=(2n+1)r_{n-1}
\label{YT}
\ee
The inversion formula corresponding to (\ref{YC}) is $P=ik-\sum_1^{\infty}
P_j(ik)^{-j}$ and one can write
\be
\partial'_{2n+1}(P^2+q)=\partial'_{2n+1}(-k^2);\,\,\partial'_{2n+1}q
=\frac{2}{2n+1}\partial r_n=\frac{2}{2n+1}\partial_qr_nq_X=
q_Xr_{n-1}
\label{YU}
\ee
Note for example $r_0=q/2,\,\,r_1=3q^2/8,\,\,r_2=5q^3/16,\cdots$ and
$\partial'_Tq=q_Xr_0=(1/2)qq_X$ (scaling is needed in (\ref{YM}) here for
comparison).  Some further calculation gives for $P=ik-\sum_1^{\infty}
P_n(ik)^{-n}$ 
\be
P_n\sim-v_n\sim-\frac{\phi_n}{2^n};\,\,c_{2n+1}=(-1)^{n+1}\int_
{-\infty}^{\infty}P_{2n+1}(X)dX
\label{YV}
\ee
The development above actually gives a connection between inverse
scattering and the dKdV theory (cf. \cite{ch,ci,cj} for more on this).

\section{BACKGROUND ON RIEMANN SURFACES}
\renewcommand{\theequation}{3.\arabic{equation}}\setcounter{equation}{0}

We recall first some ideas on BA functions and Riemann surfaces following
\cite{aa,cc,cd,ka,kb,na}.  Given a compact Riemann surface $\Sigma_g$
of genus $g$ let $(A_i,B_i)$ be a canonical homology basis, $d\omega_j$ a 
basis of normalized holomorphic differentials ($\oint_{A_j}d\omega_i=
\delta_{ij}$), ${\cal A}(P)=(\int_{P_0}^Pd\omega_k)$ the Abel-Jacobi
map ($P_o\not= P_{\infty}\sim\infty$), and $\Theta(z)=\Theta[0](z)$
the Riemann theta function.  Let $\lambda^{-1}$ be a local coordinate
near $\infty$ with $\lambda(P_{\infty})=\infty$ and take $d\Omega_j
=d(\lambda^j+O(\lambda^{-1}))$ to be normalized meromorphic differentials
of the second kind ($\oint_{A_j}d\Omega_i=0$).  Other normalizations are
also used (e.g. $\Re\oint_{A_i}d\Omega_j=\Re\oint_{B_i}d\Omega_j=0$)
but we will not dwell on this.  We set also $\Omega_{jk}=\oint_{B_k}
d\Omega_j$.  Now let $D=P_1+\cdots+P_g$ be a nonspecial divisor of degree
$g$ and set $z_0=-K-{\cal A}(D)$ where $K\sim (K_j)$ corresponds to
Riemann constants.  One can now introduce ``time" coordinates $t_j$ via
a uniquely defined BA function (up to normalization)
\be
\psi = exp(\int^P_{P_0}\sum t_nd\Omega_n)
\cdot\frac{\Theta({\cal A}(P) + \sum (t_j/2\pi i)(\Omega_{jk})+z_0)}
{\theta({\cal A}(P)+z_0)}
\label{A}
\ee
(see \cite{cc,cd} for an extensive discussion - we are working here in
in a KP framework for convenience).  Next one defines a dual divisor
$D^*$ via $D+D^*-2P_{\infty}\sim K_{\Sigma}$ where $K_{\Sigma}$
is the canonical class of $\Sigma_g$ (class of meromorphic differentials).
Then the dual BA function is (up to normalization)
\be
\psi^*\sim e^{-\int_{P_o}^P\sum t_nd\Omega_n}\cdot \frac{\Theta
({\cal A}(P)-\sum (t_j/2\pi i)(\Omega_{jk})+z_0^*)}{\Theta({\cal A}(P)
+z_0^*)}
\label{B}
\ee
($z_0^*=-{\cal A}(D^*)-K$)
and the BA conjugate differential is $(\clubsuit)
\,\,\psi^{\dagger}=
\psi^*d\hat{\Omega}$ where ($E\sim$ prime form)
\be
d\hat{\Omega}(P')=\frac{\Theta({\cal A}(P')+z_0)\Theta({\cal A}(P')+z_0^*)}
{E(P,P_{\infty})^2}
\label{C}
\ee
Thus $d\hat{\Omega}$ has zero divisor $D+D^*$ and a unique double pole
at $P_{\infty}$ so that $\psi\psi^*d\hat{\Omega}=\psi\psi^{\dagger}$
is meromorphic with a second order pole at $P_{\infty}$ and no other poles.
Note here in (\ref{A}) for example there should be a normalization factor
$c(t)$ multiplying the right side (cf. \cite{db}); we will incorporate
the normalizations via theta functions in the calculations below.
\\[3mm]\indent
It is instructive and useful to enlarge the context in the spirit of
\cite{cc,cg,cn,ia,na}.  We stay in a KP framework and write
(normalizations are now included)
\be
\psi=exp\left[\sum_1^{\infty}t_j\left(\int_{P_0}^Pd\Omega^j+\Omega^j(P_0)
\right)+i\sum_1^g
\alpha_j\left(\int_{P_0}^Pd\omega_j+\omega_j(P_0)\right)\right]\times
\label{D}
\ee
$$\times\frac{\Theta\left({\cal A}(P)+\sum_1^{\infty}(t_j/
2\pi i)(\Omega_{jk})
+i\sum_1^g\alpha_j(B_{jk})+z_0\right)\Theta({\cal A}(P_{\infty})
+z_0)}{\Theta\left({\cal A}(P_{\infty})+\sum_1^{\infty}
(t_j/2\pi i)(\Omega_{jk})
+i\sum_1^g\alpha_j(B_{jk})+z_0\right)\Theta({\cal A}(P)+z_0)}$$
(note $\int^Pd\Omega_j\sim\int_{P_0}^Pd\Omega_j
+\Omega_j(P_0)$) and ${\cal A}(P)=(\int_{P_{\infty}}^Pd\omega_j)+{\cal A}
(P_{\infty})$) and explicitly now ($z=\lambda^{-1}$ amd $q_{mj}=q_{jm}$)
\be
d\Omega_j=d\Omega^j\sim d\left(\lambda^j-\sum_1^{\infty}\frac{q_{mj}}{m}
z^m\right);\,\,d\omega_j\sim d\left(-\sum_1^{\infty}
\sigma_{jm}\frac{z^m}{m}\right);\,\,\Omega_{nj}=2\pi i\sigma_{jn}
\label{E}
\ee
(see \cite{cc} for details).  There is also a general theory of
prepotential etc. following \cite{cc,ia,na} for example which involves
($T_n\sim\epsilon t_n$ as indicated below)
\be
d{\cal S}=\sum_1^ga_jd\omega_j+\sum_1^{\infty}T_nd\Omega_n;\,\,
\frac{\partial d{\cal S}}{\partial a_j}=d\omega_j;\,\,\frac{\partial d{\cal S}}
{\partial T_n}=d\Omega_n
\label{F}
\ee
If we consider functions $F(a,T)$ related to $d{\cal S}$ via
\be
\frac{\partial F}{\partial a_j}=\frac{1}{2\pi i}\oint_{B_j}d{\cal S};\,\,
\partial_nF=-Res_{\infty}z^{-n}d{\cal S}
\label{G}
\ee
then, given the standard 
class of solutions of the Whitham hierarchy satisfying (cf. \cite{cc,kc})
\be
2F=\sum_1^ga_j\frac{\partial F}{\partial a_j}+\sum_1^{\infty}T_n\frac
{\partial F}{\partial T_n}
\label{H}
\ee
there results
\be
2F=\sum_1^g\frac{a_j}{2\pi i}\oint_{B_j}d{\cal S}-\sum_1^{\infty}
T_nRes_{\infty}z^{-n}d{\cal S}
\label{I}
\ee
Writing now, in the notation of \cite{na},
$d\omega_j=-\sum_1^{\infty}\sigma_{jm}z^{m-1}dz$ with $d\Omega_n=
[-nz^{-n-1}-\sum_1^{\infty}q_{mn}z^{m-1}]dz$, and using (\ref{F}),
one obtains ($B_{jk}$ is the period matrix)
\be
2F=\frac{1}{2\pi i}\sum_{j,k=1}^gB_{jk}a_ja_k+2\sum_1^ga_j\sum_1^{\infty}
\sigma_{jk}T_k+\sum_{k,l=1}^{\infty}q_{kl}T_kT_l
\label{J}
\ee
\indent
Thus the expression (\ref{J}) comes from the Riemann surface theory, 
without explicit reference to the BA function, and we
consider now (\ref{D}) and 
\be
\psi=exp\left(\sum_1^{\infty}t_i\lambda^i\right)\times\frac{\tau
(t-[\lambda^{-1}],\alpha)}{\tau(t,\alpha)}
\label{K}
\ee
to which ideas of dKP can be applied to 
introduce the slow variables $T_k$.  This means that we will
be able to introduce
slow variables in two different ways and the resulting comparisons will show
an equivalence of procedures.  In practice this will enable one to treat
$\epsilon$ on the same footing in the Whitham theory and in the dispersionless
theory
(see also \cite{cd} for an approach based on \cite{aa}).
Thus from (\ref{D}) and (\ref{K}) one obtains 
an expression for $\tau$ of the form
($t_1=x,\,\,t_2=y,\,\,t_3=t,\,\cdots$)
\be
\tau(t,\alpha)=exp[\hat{F}(\alpha,t)]\Theta\left(
{\cal A}(P_{\infty})+\sum_1^{\infty}(t_j/2\pi i)(\Omega_{jk})
+i\sum_1^g\alpha_j(B_{jk})+z_0\right)
\label{L}
\ee
where $k=1,\cdots,g$ and
\be
\hat{F}(\alpha,t)=\frac{1}{2}\sum_{k,l=1}^{\infty}q_{kl}t_kt_l-\frac
{1}{4\pi i}\sum_{j,k=1}^{\infty}B_{jk}\alpha_j\alpha_k+i\sum_1^g
\alpha_j\sum_1^{\infty}\sigma_{jk}t_k+\sum_1^{\infty}d_kt_k
\label{M}
\ee
(see also \cite{kd} for a similar form - recall here ${\cal A}(P)=
(\int_{P_0}^Pd\omega_j)$ and $P_0\not= P_{\infty}$ is required).
Putting in the slow variables $T_k=\epsilon t_k$ and $a_k=i\epsilon\alpha_k$
one will find
that the quadratic part of $\hat{F}(T/\epsilon,a/i\epsilon)$ 
in $T$ and $a$ is exactly $F(a,T)/\epsilon^2$ for $F$ in (\ref{J});
here $\tau=exp[(1/\epsilon^2)\tilde{F}+O(1/\epsilon)]$
(with $\tilde{F}/\epsilon^2$ the quadratic part of $\hat{F}(T/\epsilon,
a/i\epsilon)$)
is the natural form of $\tau$ based
on (\ref{K}) and it is associated with $\psi\sim exp[(1/\epsilon)
S+O(1)]$ (cf. \cite{ci} - note this is $S$ and not ${\cal S}$ - $S$
will be discussed later in Section 4).
In \cite{na} one writes then from (\ref{L}) and (\ref{D})
respectively 
\be
\frac{1}{\epsilon^2}F(a,T)+O\left(\frac{1}{\epsilon}\right)=
log\tau\left(\frac{T}{\epsilon},\frac{a}{i\epsilon}\right)=\epsilon^{-2}
\sum_0^{\infty}\epsilon^nF^{(n)}(T,a);
\label{N}
\ee
$$dlog\psi\left(p,\frac{T}{\epsilon},
\frac{a}{i\epsilon}\right)=\epsilon^{-1}\sum_0^{\infty}\epsilon^n
d{\cal S}^{(n)}(p,T,a)\sim d\left(\frac{1}{\epsilon}S+O(1)\right)$$
where $d{\cal S}^{(0)}\sim d{\cal S}$ in (\ref{F}) and $F^{(0)}\sim F$
in (\ref{J}).  Suitable calculations are displayed in \cite{cc} to
establish the relations between $F$ and $\hat{F}$ as indicated.
\\[3mm]\indent
For perspective however let us make now a few background observations.  First
we refer first to \cite{ch} where it
is proved that $F_{mn}=F_{nm}$ in ${\cal B}_n=\lambda^n-\sum_1^{\infty}
(F_{nm}/m)\lambda^{-m}$ (the $F_{mn}$ being treated as algebraic symbols
with two indices generally and $F_{mn}=\partial_m\partial_nF$
specifically).
Since near the point at infinity we have $\Omega_n\sim
\lambda^n-\sum_1^{\infty}(q_{mn}/m)\lambda^{-m}$ the same sort of proof
by residues is suggested ($F_{mn}=-Res_{\lambda}[{\cal B}_nd\lambda^m]$)
but we recall that ${\cal B}_n=\lambda^n_{+}$ so there is an underlying
$\lambda$ for all ${\cal B}_n$ which makes the proof possible.  Here 
one should be careful however.  For example 
$(\spadesuit)\,\,
p=\lambda-\sum_1^{\infty}(H_j/j)\lambda^{-j}$ corresponds to $P=\lambda
+\sum_1^{\infty}P_{j+1}\lambda^{-j}$ in \cite{ch} with $P_{j+1}=F_{1j}/j$
(i.e. $H_j\sim -F_{1j}$) and the ``inverse" is $\lambda=P+\sum_1^{\infty}
U_{n+1}P^{-n}$ (arising from a Lax operator $L$ via dKP).  The corresponding
inverse for $(\spadesuit)$ 
then characterizes $\lambda$ in terms
of $p$ but one does not automatically 
expect $\Omega_n\sim\lambda^n_{+}$.  The matter is somewhat
subtle.  Indeed the BA function is defined from the Riemann surface via
$d\Omega_n,\,\,d\omega_j$, and normalizations.  It then produces a unique
asymptotic expansion at $\infty$ which characterizes $\psi$ near $\infty$ in
terms of $\lambda$ and hence must characterize the $d\Omega_n$ and $d
\omega_j$ asymptotically.  Moreover the normalizations must be built into
these expansions since they were used in determining $\psi$.  
Thus we must have $F_{mn}\sim q_{mn}$ as a consequence of the
BA function linking the differentials and the asymptotic expansions
(note also that the 
formal algebraic determination of ${\cal B}_n$ via $\lambda^n_{+}$
is a consequence of relating the $d\Omega_n$ to operators $L_n=
L^n_{+}$ as in \cite{ka} which corresponds to looking at $\lambda^n_{+}$
with $\lambda=P+\sum_1^{\infty}U_{n+1}P^{-n}$ as above). 
Another approach
(following \cite{cd}) is to extract from remarks after (\ref{M}) that
$q_{mn}=F_{mn}$ at $T^0_k=0$ via $F_{mn}=\partial_m\partial_nF$,
so that expanding around an arbitrary
$T_k^0$ as in \cite{kc} one can assert that $q_{mn}=F_{mn}$ with 
arbitrary argument.  
Even better is to identify $dS$ and $d{\cal S}$ via uniqueness of the
BA function and then derive $\partial_ndS=d{\cal B}_n=\partial_n
d{\cal S}=d\Omega_n$.
Further
with this identification we recover the Whitham equations as in \cite{cd}
via
\be
\partial_k\Omega_n=-\sum_{m,n=1}^{\infty}\frac{F_{mnk}}{m}z^m=
\partial_n\Omega_k=-\sum_{m,n=1}^{\infty}\frac{F_{mkn}}{m}z^m
\label{O}
\ee
Finally we recall now that in SW duality one sets $a_j^D=\partial F/
\partial a_j$ and the formulas (\ref{F}) - (\ref{I}) are fundamental
relations (see e.g. \cite{cc,cg,da,ia,ka,na}).
Note also that our Riemann surface will be eventually based on KdV
situations so it will be hyperelliptic and can be viewed as a branched
surface with transcendentality $\prod_1^{2g+1}(\lambda-\lambda_j)$ 
with $\infty$ also a branch point (here $\lambda_j\in {\bf R},\,\,
\lambda_1<\lambda_2<\cdots$ - cf. \cite{cc}). 

\section{BACKGROUND ON $(X,\psi)$ DUALITY}
\renewcommand{\theequation}{4.\arabic{equation}}\setcounter{equation}{0}

We extract first from \cite{ce} to indicate the duality of \cite{fa}
between $X$ and $\psi$ (cf. also \cite{ba,ca,cb}).
The point of departure is the Schr\"odinger equation
\be
{\cal H}\psi_E=-\frac{\hbar^2}{2m}\psi''_E+V(X)\psi_E=E\psi_E
\label{AA}
\ee
where $X$ is the quantum mechanical (QM) space variable with $\psi'_E=
\partial\psi_E/\partial X$ and we write $\epsilon=\hbar/\sqrt{2m}\,\,
(E$ is assumed real).  
The theme of \cite{fa} is very important, perhaps paradigmatic, and
is developed further in \cite{fy}; we suspect there are significant
connections to \cite{oa} as well (cf. Remark 6.11).
In \cite{ca,cb} we discussed the possible origin of
this from a Kadomtsev-Petviashvili (KP) situation $L^2_{+}\psi=\partial\psi/
\partial t_2$ where $L^2_{+}=\partial_x^2-v(x,t_i)$ and e.g. $\tau_2=
-i\sqrt{2m}T_2$ so $\partial_{t_2}=\epsilon\partial_{T_2}=-i\hbar
\partial_{\tau_2}$ (one writes $X=\epsilon x$ and 
$T_i=\epsilon t_i$ in the dispersionless theory - in the enlarged
context of (\ref{D}) on a Riemann surface one supplements this with 
$a_k=i\epsilon\alpha_k$, but we will suppress the $\alpha_k$ here for
convenience).
This leads to an
approximation
\be
\epsilon^2\psi''_E-V(X,T_i)\psi_E\sim\epsilon\frac{\partial\psi_E}{\partial T_2}
=-i\hbar\frac{\psi_E}{\partial\tau_2} 
\label{BB}
\ee
corresponding to the Schr\"odinger equation.  This is also related to the
Korteweg-deVries (KdV) equation and it's dispersionless form dKdV as
indicated below (cf. Section 5 for more discussion of (\ref{AA}) - (\ref{BB})).
\\[3mm]\indent {\bf REMARK 4.1}$\,\,$  
For the approximation of potentials one
assumes e.g. $v=v(x,t_i)\to v(X/\epsilon,T_i/\epsilon)
=V(X,T_i)
+O(\epsilon)$.  This is
standard in dispersionless KP = dKP and certainly realizable
by quotients of homogeneous polynomials for example. 
In fact it is hardly a restriction since given e.g. $F(X)=\sum_0^{\infty}
a_nX^n$ consider $\tilde{f}(x,t_i)=a_0+\sum_1^{\infty}(x^n/\prod_2^{n+1}
t_i)$.  Then $\tilde{f}(X/\epsilon,T_i/\epsilon)=a_0+\sum_1^{\infty}
(X^n/\prod_2^{n+1}T_i)$ and one can choose the $T_i$ recursively so that
$1/T_1=a_1,\,\,1/T_1T_2=a_2,\cdots$, leading to $F(X)=\tilde{F}(X,T_i)$. 
\\[3mm]\indent {\bf REMARK 4.2.}$\,\,$ Returning to (\ref{BB}),
when
$\psi_E=exp(S/\epsilon)$ for example, one has $\epsilon\psi'_E=S_X\psi_E$
with $\epsilon^2\psi''_E=\epsilon S_{XX}\psi_E+(S_X)^2\psi_E$ so in (\ref{BB})
we are neglecting an $O(\epsilon)\psi_E$ term from $v$, and for $\psi_E
=exp(S/\epsilon)$ another $\epsilon S_{XX}\psi_E$ term is normally removed
in dispersionless theory.  Then for ${\cal H}$ independent of $\tau_2$ for
example one could assume $V$ is independent of $T_2$ and write formally
in (\ref{BB}), $\hat{\psi}_E=exp(E\tau_2/i\hbar)\cdot\psi_E$, 
with ${\cal H}\psi_E
=E\psi_E$, which is (\ref{AA}).  Since in the QM problem one does not
however run $\hbar\to 0$ (hence $\epsilon\not\to 0$) one should argue
that these $O(\epsilon)$ terms should be retained,
and we will develop this approach, which essentially corresponds to WKB
(with some background structure).
In particular one could ask
for $v(X/\epsilon,T_i/\epsilon)=V(X,T_i)+\epsilon\hat{V}
(X,T_i)+O(\epsilon^2)$ and retain the 
$\epsilon\hat{V}$ term along with $\epsilon
S_{XX}$, in requiring e.g. $S_{XX}=\hat{V}$ (this is covered 
below - an additional term also arises).
In fact, to establish a connection with quantum mechanics and the 
Schr\"odinger equation,
the passage from $v\to V$ or $V+\epsilon \hat{V}$ is the only
``assumption" in our development below and this admits various realizations;
the impact here only involves some possible minor restrictions on
the class of quantum potentials
to which the theory applies.  The background mathematics behind
$V$ determined by KP or KdV essentially generates some additional structure
which allows us to insert $X$ into the theory in a manner commensurate
with its role in \cite{fa}.  The formulation of \cite{fa} then entails
some constraints on the background objects as indicated in the text.
We emphasize that inserting $S$ is familiar from WKB (cf. \cite{mb,mc});
we are introducing in an ad hoc manner additional
variables $T_i$ or $T_i,\,\,\lambda$ or $k$, etc. to spawn a KP or KdV
theory.  We do not assume or even suggest that this is in any way connected
a priori with the physics of the quantum mechanical problem (although
of course it conceivably could be since integrability ideas are important
in quantum mechanics).  This procedure generates a nice Hamilton-Jacobi (HJ)
theory which guides one to insert $X$ into the machinery, but the
insertion itself is at ``ground level" and simply reflects a WKB
formulation; neither the underlying KdV or KP dynamics nor the HJ
theory is directly used here.  Once $X$ is involved connections to
\cite{fa} are immediate.  Actually the procedure could be reversed
as a way of introducing duality ideas into the $\epsilon$-dispersionless
theory of \cite{cb,ce}) and this should probably be
related to the duality already studied in Whitham theory (cf. 
\cite{cc,cg,ia,ka,ma}), given a finite zone theory on a Riemann surface.
Thus start with KdV or KP, go to the Schr\"odinger
equation and $dKdV_{\epsilon}$ or $dKP_{\epsilon}$, develop the HJ theory,
and then use \cite{fa} to create duality.  More generally, start from
a finite zone KdV situation with associated Whitham dynamics on a
Riemann surface and compare dualities; this is the aim of the
present paper.
\\[3mm]\indent
We list first a few of the equations from \cite{fa}, as written in 
\cite{ca,cb,ce}, without a discussion
of philosophy (some of which will
be mentioned later).  Thus ${\cal F}$ is a prepotential
and, since $E$ is real,
$\psi_E$ and $\bar{\psi}_E=\psi_E^D$ both satisfy (\ref{AA}) with $\psi_E^D=
\partial{\cal F}/\partial\psi_E$.  
The Wronskian in (\ref{AA}) is taken to be
$W=\psi'\bar{\psi}-\psi\bar{\psi}'=2\sqrt{2m}/i\hbar=2/i\epsilon$ and one
has ($\psi=\psi(X)$ and $X=X(\psi)$ with $X_{\psi}=\partial X/\partial\psi=
1/\psi'$)
\be
{\cal F}'=\psi'\bar{\psi};\,\,{\cal F}=\frac{1}{2}\psi\bar{\psi}+\frac
{X}{i\epsilon};\,\,
\frac{\partial\bar{\psi}}{\partial\psi}=\frac{1}{\psi}\left[\bar{\psi}-
\frac{2}{i\epsilon}X_{\psi}\right]
\label{FF}
\ee
($\psi$ always means $\psi_E$ but we omit the subscript occasionally
for brevity).
Setting $\phi=\partial{\cal F}/\partial(\psi^2)=\bar{\psi}/2\psi$ with
$\partial_{\psi}=2\psi\partial/\partial(\psi^2)$ and evidently
$\partial\phi/\partial
\psi=-(\bar{\psi}/2\psi^2)+(1/2\psi)(\partial\bar{\psi}/\partial\psi)$
one has a Legendre transform pair
\be
-\frac{X}{i\epsilon}=\psi^2\frac{\partial{\cal F}}{\partial(\psi^2)}
-{\cal F};\,\,-{\cal F}=\phi\frac{1}{i\epsilon}X_{\phi}-\frac{X}{i\epsilon}
\label{HH}
\ee
One obtains also $(\spadesuit\bullet\spadesuit)\,\,
|\psi|^2=2{\cal F}-(2X/i\epsilon)\,\,({\cal F}_{\psi}=\bar{\psi});\,\,
-(1/i\epsilon)X_{\phi}=\psi^2;\,\,{\cal F}_{\psi\psi}=
\partial\bar{\psi}/\partial\psi$.
Further from $X_{\psi}\psi'=1$ one has $X_{\psi\psi}\psi'+X^2_{\psi}\psi''
=0$ which implies
\be
X_{\psi\psi}=-\frac{\psi''}{(\psi')^3}=\frac{1}{\epsilon^2}\frac
{(E-V)\psi}{(\psi')^3}
\label{KK}
\ee
\be
{\cal F}_{\psi\psi\psi}=\frac{E-V}{4}({\cal F}_{\psi}-\psi\partial^2_{\psi}
{\cal F})^3=\frac{E-V}{4}\left(\frac{2X_{\psi}}{i\epsilon}\right)^3
\label{LL}
\ee
Although a direct comparison of (\ref{LL}) to the Gelfand-Dickey resolvant
equation ($(\clubsuit\clubsuit)$ below) 
is not evident ($V'$ is lacking) a result of
T. Montroy which expands ${\cal F}_{\psi\psi\psi}$
shows that in fact (\ref{LL}) corresponds exactly to
\be
\epsilon^2{\cal F}'''+4(E-V)\left({\cal F}'-\frac{1}{i\epsilon}\right)
-2V'\left({\cal F}-\frac{X}{i\epsilon}\right)=0
\label{ZZZ}
\ee
which is $(\clubsuit\clubsuit)$ since $\Xi=|\psi|^2=2{\cal F}-(2X/i\epsilon)$.
\\[3mm]\indent
Next there is a so-called eikonal transformation (cf. \cite{mf}) which can
be related to \cite{fa} as in \cite{ca,cb,ce}.  We consider
real $A$ and $S$ with
$\psi=Ae^{(i/\hbar)S};\,\,p=ASin[(1/\hbar)S];\,\,q=ACos[(1/\hbar)
S]$.  Then
introducing new variables 
$\chi=A^2=|\psi|^2;\,\,\xi=(1/2\hbar)S$
it follows that there will be a Hamiltonian format with symplectic form
$(\spadesuit\spadesuit)\,\,\delta p\wedge
\delta q=\delta\xi\wedge\delta\chi=\tilde{\omega}$.
It is interesting to write down the connection between the $(S,A)$ or
$(\chi,\xi)$ type
variables and the variables from \cite{fa} and it will be useful to  
take now $\psi=Aexp(iS/\epsilon)\,\,(\epsilon=\hbar/\sqrt{2m})$ with
$\xi\sim S/2\epsilon$. Then
\be
{\cal F}=\frac{1}{2}\chi+\frac{X}{i\epsilon};\,\,{\cal F}'=
\psi'\bar{\psi} =\frac{1}{2}\chi'
+\frac{i}{\epsilon}P\chi
\label{TT}
\ee
for $S'=S_X=P$ and there is an interesting relation
$(\clubsuit\bullet\clubsuit)\,\,
P\chi=-1\Rightarrow \delta\chi=-(\chi/P)\delta P$.
Further from $\phi=(1/2)exp[-(2i/\epsilon)S]$ and $\psi^2=\chi exp(4i\xi)$ 
we have
\be
\psi^2\phi=\frac{1}{2}\chi=-\frac{1}{\epsilon}\phi X_{\phi};\,\,\xi=\frac
{S}{2\epsilon}=\frac{i}{4}log(2\phi)
\label{VV}
\ee
Now the theory of the Seiberg-Witten (SW) differential $\lambda_{SW}$ 
following \cite{ca,cc,cg,da,ia,ka,ma} for example 
involves finding a differential
$\lambda_{SW}$ of the form $QdE$ or $td\omega_0$ (in the spirit of
\cite{ka} or \cite{da,ia} respectively) such that $d\lambda_{SW}=\omega$
is a symplectic form.  In the
present context one can ask now whether the form $\tilde{\omega}$ of 
$(\spadesuit\spadesuit)$ makes any sense 
in such a context.  Evidently this is jumping
the gun since there is no Riemann surface in sight (see however \cite{ca}
for a Riemann surface with some validation 
and variation as in \cite{cc,cd,ck} - this
is developed below in certain directions).
Some motivation to consider 
the matter here comes from the
following formulas which express $\tilde{\omega}$ nicely in terms of
the duality variables of \cite{fa} (another version of a ``canonical"
symplectic form in terms of ${\cal F}$ alone is given below).
Thus a priori $\psi=\Re\psi+i\Im\psi$ has two components which are also
visible in $\psi=Aexp(iS/\epsilon)$ as $A$ and $S$.  The relation $P\chi
=\chi(\partial S/\partial X)=-1$ indicates a dependence between $A$
and $S'$ (but not $A$ and $S$) which is a consequence of the duality between
$\psi$ and $X$.  Then $2AS'\delta A+A^2\delta S'=0$ or $\delta S'
=-(2S'/A)\delta A\equiv (\delta S'/S')=-2(\delta A/A)$, whereas $\delta
\psi/\psi\sim 2(\delta A/A) +(i/\epsilon)\delta S$.  It follows that
$\Re(\delta\psi/\psi)=-(\delta S'/S')$ and $\Im(\delta \psi/\psi)=(\delta
S/\epsilon)$.  The sensible thing seems to be to look at the complex
dependence of $X(\psi)$ and $\psi(X)$ in terms of two real variables
and $\delta\xi\wedge\delta\chi$ will have a nice
form in transforming to the variables of \cite{fa}.  In particular
from $\psi^2\phi=(1/2)\chi$ with $\delta\chi=4\phi\psi\delta\psi+
2\psi^2\delta\phi$ we obtain
$(\delta\psi/\psi)=2(\delta\chi/\chi)-
(\delta\phi/\phi)$.
Hence one can write 
\\[3mm]\indent {\bf PROPOSITION 4.3.}$\,\,$ Under the hypotheses indicated
\be
\delta\xi\wedge\delta\chi=\frac{i}{4}\frac{\delta\phi}{\phi}
\wedge\chi\frac{\delta\chi}{\chi}=\frac{i}{2}\delta\phi\wedge\delta\psi^2
\sim\frac{i}{2}\delta\bar{\psi}\wedge\delta\psi
\label{XX}
\ee
(note $\delta\phi=(1/2\phi)\delta\bar{\psi}-(\bar{\psi}/2\psi^2)\delta\psi$)
and in an exploratory spirit the differentials $\lambda=(i/2)\phi\delta
\psi^2$ or $\lambda=(i/2)\psi^2\delta\phi$, along
with $\lambda=(i/2)\bar{\psi}\delta\psi$
or $\lambda=(i/2)\psi\delta\bar{\psi}$, might merit further consideration.
\\[3mm]\indent
We refer now to \cite{ch,ci,cj,ta} for dispersionless KP ($=$ dKP) and consider
here $\psi=exp[(1/\epsilon)S(X,T,\lambda)]$ instead of $\psi=Aexp(S/\epsilon)$
(more details are given later).
Thus $P=S'=S_X$ and $P^2=V-E$ 
but $E\not= \pm\lambda^2$ (unless otherwise stated) and
this does not define $S$ via $P=S_X$ unless we have a KdV situation
(which does not
seem a priori desirable but in fact will be seen to be the natural format here
upon development with modifications
of the dispersionless theory - cf. \cite{cb});  
thus generally $\lambda$ is the $\lambda$ of $S(T_n,\lambda)$ from KP theory
and we recall that $\psi$ always means $\psi_E$ as in \cite{fa}.
Some routine calculation yields
(recall $X_{\psi}=1/\psi'$ and $\psi'=(P/\epsilon)\psi$)
\be
\phi=\frac{1}{2}e^{-(2i/\epsilon)\Im S};\,\,\frac{1}{\epsilon}X_{\phi}=
-ie^{(2/\epsilon)S};\,\,X_{\psi}=\frac{\epsilon}{P}e^{-S/\epsilon}
\label{AAQ}
\ee
\be
\frac{1}{\epsilon}X_{\psi\psi}=\frac{E-V}{P^3}e^{-S/\epsilon};\,\,
{\cal F}_{\psi}=\bar{\psi}=e^{\bar{S}/\epsilon};\,\,
{\cal F}_{\psi\psi}=e^{-(2i/\epsilon)\Im S}-\frac{2}{iP}e^{-2S/\epsilon}
\label{ABQ}
\ee
\be
|\psi|^2=e^{(2/\epsilon)\Re S};\,\,\frac{S}{\epsilon}=\frac{1}{2}log|\psi|^2
-\frac{1}{2}log(2\phi);\,\,\bar{P}=\bar{S}_X=P-\frac{2}{i\psi\bar{\psi}}
\label{ACQ}
\ee
Summarizing one has
\be
\Im{\cal F}=-\frac{X}{\epsilon};\,\,\Re{\cal F}=\frac{1}{2}|\psi|^2=
\frac{1}{2}e^{\frac{2}{\epsilon}\Re S}=-\frac{1}{2\Im P}
\label{AGQ}
\ee
In the present situation $|\psi|^2=exp[(2/\epsilon)\Re S]$ and $2\phi=
exp[-(2i/\epsilon)\Im S]$ can play the roles of independent variables
(cf. (\ref{ACQ}).  The version here of $P\chi=-1$ is $\chi\Im P=-1$,
while $\psi^2\phi=(1/2)|\psi|^2=(1/2)\chi$ again, and
we obtain as above the formula (\ref{XX}). 
Now note that for $L=\partial+\sum_1^{\infty}u_i
\partial^{-i},\,\,L^2_{+}=\partial^2+2u_1$, and $u_1=\partial^2log(\tau)$
where $\tau$ is the famous tau function.
This implies $v=-2\partial^2log(\tau)$ here, from which $V=-2F_{XX}$ for
$\tau=exp[(1/\epsilon^2)F+O(1/\epsilon)]$ 
in the dispersionless theory (cf. \ref{N})).  We recall also 
the Gelfand - Dickey
resolvant equation (cf. \cite{co}) for $\Xi=\psi\bar{\psi}$, namely,
in the present notation 
$(\clubsuit\clubsuit)\,\,
\epsilon^2\Xi'''-4V\Xi'-2V'\Xi+4E\Xi'=0$
(direct calculation).  
Using $\Xi=
2{\cal F}-(2X/i\epsilon),\,\,\Xi'=2{\cal F}'-(2/i\epsilon),\,\,\Xi''=
2{\cal F}''$, and $\Xi'''=2{\cal F}'''$, we obtain 
then from $(\clubsuit\clubsuit)$ (cf. also (\ref{ZZZ}))
\be
\epsilon^2{\cal F}'''+\left({\cal F}'-\frac{1}{i\epsilon}\right)(8F''
+4E)+4F'''\left({\cal F}-\frac{X}{i\epsilon}\right)=0
\label{AKQ}
\ee
which provides a relation between $F$ and ${\cal F}$.
We will see below how to embellish all this with a new modification of
the dKP and dKdV theory.
Thus we state here heuristically
\\[3mm]\indent {\bf THEOREM 4.4.}$\,\,$  Under the hypotheses indicated
of first order WKB type approximation,
the equation (\ref{AKQ}) yields a relation between the prepotential
${\cal F}$ of $(X,\psi)$ duality defined via (\ref{FF}) and the prepotential
$F(a,T)$ of (\ref{J}) (also corresponding to a free energy in dKP or dKdV).
\\[3mm]\indent 
The exposition to follow using an
expanded $dKdV_{\epsilon}$ theory based on \cite{cb} will establish
more refined relations.
One also sees that the Riemann surface background produces the $a_i$ variables
naturally here and we want now to find a definition of ${\cal F}$ which is
based on dKdV quantities and not on $\psi$ directly.  Perhaps this will
suggest another way to view duality based on ${\cal F}$.  One notes that
the word duality involving ${\cal F}$ refers to $X$ and $\psi$ whereas 
duality in SW theory refers to $a_i$ and $a_i^D=\partial F/\partial a_i$
as being dual.  In ${\cal F}$ of (\ref{FF}) of course $\bar{\psi}=\psi^D=
\partial {\cal F}/\partial\psi$ but it is $X$ and $\psi$ which are said to
be dual.  It will be shown below (following \cite{cb,ce}) that
$dX\wedge dP\sim [-\epsilon/2(\Re{\cal F})^2]d\Im({\cal F})\wedge
d(\Re{\cal F})$ follows from the first order WKB aspects
of dKP where $P=S_X$.  On the othe hand, following \cite{ca,cc,ka}, one has
a canonical symplectic form $\omega\sim\sum da_i\wedge d\omega_i$ associated
with SW theory.  A priori there seems to be no conceptual reason why SW
theory should have any relation to $(X,\psi)$ duality, except perhaps 
that the background mathematics and development in \cite{ba,ma} has many
features related to SW mathematics.  The connection indicated by (\ref{AKQ})
relating $F(a,T)$ and ${\cal F}$ is momentarily purely formal; it may
not signify much in terms of conceptual meaning and this will be
pursued below.  The natural occurance of a symplectic form
$(i/2)\delta\bar{\psi}\wedge \delta\psi$ in (\ref{XX}) suggests a ``duality"
analogue involving $\sum da_i^D\wedge da_i$ but there seems to
be no immediate conceptual connection here.
In any event, although $\epsilon$ (or $\hbar$) seems to dangle in the
formulas of this section, one is accustomed to this is quantum mechanics
and here it can be regarded as a scale parameter (cf. \cite{fa}).
Below, in modifying the dKP or dKdV theory to $dKdV_{\epsilon}$ for example
we will balance powers of $\epsilon$.

\section{REFINEMENTS FOR dKP and dKdV}
\renewcommand{\theequation}{5.\arabic{equation}}\setcounter{equation}{0}

Now
the dKP theory as in \cite{ch,ci,ke,ta} involves a parameter $\epsilon\to
0$ and we recall $L=\partial+\sum_1^{\infty}u_{n+1}(t)\partial^{-n}\to
L_{\epsilon}=\epsilon\partial +\sum_1^{\infty}u_{n+1}(\epsilon,T)(\epsilon
\partial)^{-n}$ where $t\sim (t_k),\,\,T\sim (T_k),$ and $X=T_1$ with
$u_{n+1}(\epsilon,T)=U_{n+1}(T)+O(\epsilon)$ as in Section 2.  Then for
$\psi=exp(S/\epsilon)$ one has $L\psi=\lambda\psi\to\lambda= P+\sum_1^{\infty}
U_{n+1}P^{-n}$ where $P=S_X$ with $S=S(X,T_k,\lambda)\,\,(k\geq 2)$.
Here all the terms which are $O(\epsilon)$ are passed to zero but
$\epsilon\not\to 0$ in the QM situation 
related to \cite{fa} where $\epsilon=\hbar/\sqrt
{2m}$ and we want to develop further the first order connections
to \cite{fa} indicated above and somehow balance the $\epsilon$ terms.
Hence one
thinks of rewriting some of the dKP theory for example in order to retain
$O(\epsilon)$ terms at least and we have referred to this as
$dKP_{\epsilon}$ theory; it essentially corresponds
to an expanded WKB with the proviso 
that there is a background mathematics
providing some additional structure (details are given below).
In this direction we recall that 
$S=\sum_1^{\infty}T_n\lambda^n+\sum_1^{\infty}S_{j+1}\lambda^{-j}$ and 
${\cal B}_n=\partial_nS=\lambda^n+\sum_1^{\infty}\partial_nS_{j+1}
\lambda^{-j}$.
Hence via $log\psi=(S/\epsilon)+O(1)\sim log\tau[\epsilon,T_n-(\epsilon/
n\lambda^n)]-log\tau+\sum_1^{\infty}T_n\lambda^n/\epsilon$ with $log\tau
=(F/\epsilon^2)+O(1/\epsilon)$ and $F[T_n-(\epsilon/n\lambda^n)]-F(T_n)\sim
-\epsilon\sum_1^{\infty}(\partial_nF/n\lambda^n)+O(\epsilon^2)$ one 
obtains $S_{n+1}=-(\partial_nF/n)$.  Consider now the next order terms
via ($F$ is real and $a_k$ is not involved) 
\be
F\left(T_n-\frac{\epsilon}{n\lambda^n}\right)-F(T_n)=-\epsilon\sum_1^{\infty}
\left(\frac{\partial_nF}{n\lambda^n}\right)+\frac{\epsilon^2}{2}\sum
\left(\frac{F_{mn}}{mn}\right)\lambda^{-m-n}+O(\epsilon^3)
\label{five}
\ee
Thus $\Delta log\tau=(1/\epsilon^2)\Delta F$ has $O(1)$ terms $(1/2)\sum
(F_{mn}/mn)\lambda^{-m-n}$ which correspond to the $O(1)$ terms in $log\psi$.
Hence we have a natural way of writing
$\tilde{S}=S^0+\epsilon S^1$ with $S^0=S$ and
\be
S^1=\frac{1}{2}\sum\left(\frac{F_{mn}}{nm}\right)\lambda^{-m-n};\,\,
\tilde{S}_X\sim P+\frac{\epsilon}{2}\sum\left(\frac{F_{1mn}}{nm}\right)\lambda^
{-m-n}
\label{six}
\ee
One will eventually also include $F=F^0+\epsilon F^1+\cdots$, etc. 
(cf (\ref{N}))
\\[3mm]\indent
We will carry out our discussion of enhanced KP and KdV in connection
with the background structures of Section 4 (which provide motivation).
This seems more meaningful than simply writing out terms as in
(\ref{five}) - (\ref{six}) to produce a maze of formulas.  We will only
develop $dKdV_{\epsilon}$ in detail after indicating why this is the
appropriate theory.  The techniques can be extended to $dKP_{\epsilon}$
in an obvious way.
\\[3mm]\indent {\bf REMARK 5.1.}$\,\,$  First, following \cite{ca},
we note that preliminary considerations suggest a dKP1 format.  Indeed
consider (\ref{BB}) in the form $i\hbar\psi_{\tau}=-(\hbar^2/2m)\psi''
+V\psi={\cal H}\psi$ and recall that in KP1 the even variables can be
taken as imaginary in many natural situations (cf. \cite{co}).  Hence
an equation $\psi_t=B_2\psi=\psi_{xx}-v\psi$ with $t$ imaginary is natural
and one can imagine (\ref{BB}) arising from a dispersionless KP1 situation.
An enticing possibility here is
to note that the identity
$\bar{\psi}\sim\psi^D$ in \cite{fa} could be extended to $\psi^*\sim\psi^D$
where $\psi^D$ now refers to duality in the QM sense while $\psi^*$
denotes KP duality.  Note here that for $V$ real one can
write ($B_2\sim L^2_{+}$)
\be
\partial_y\psi=(\partial_x^2-v)\psi=-H\psi=B_2\psi;\,\,-\partial_y\psi^*
=B_2^*\psi=B_2\psi^*=-H\psi^*
\label{EEE}
\ee
But for $y=it,\,\,\psi=exp(-itH)\psi_0=exp(-yH)\psi_0$ implies $\bar{\psi}=exp
(itH)\bar{\psi}_0=exp(yH)\bar{\psi}_0$ so $\bar{\psi}\sim\psi^*=exp(yH)
\psi^*_0$ for $\psi^*=\bar{\psi}_0$. This suggests that (perhaps in a limited
way only) one can relate $\psi^*$ to $\psi^D$ and envision 
the $(X,\psi)$ duality as
a special form of $\psi - \psi^*$ duality.  Generally of course $\psi^*
\not=\bar{\psi}$ and $E$ real in (\ref{AA}) is generally inappropriate for
KP, but in view of the role of $\psi,\,\,\psi^*$ in the study of
symmetries and Whitham theory for example, the use of forms $\delta\psi
\wedge\delta\psi^*$ and related objects is known to be productive
(cf. \cite{cc,cd,cg,cm,fb,ka,kb,kc,kf}) and hence the duality theme
with prepotential defined as in \cite{fa} could well have a version
involving $\psi$ and $\psi^*$ in a much more general context.
\\[3mm]\indent
In any event these considerations demand a preliminary investigation of
dKP (dKP1 in particular) and this was sketched in \cite{cb} with primary
attention to $O(\epsilon)$ terms (e.g. $\tilde{S}=S^0+\epsilon S^1+
O(\epsilon^2)$).  A first observation here tells us that if $\psi$ is to
be a quantum mechanical wave function with $|\psi|^2=exp[(2/\epsilon)
S]$ as in (\ref{ACQ}) with $|\psi|^2\leq 1$ then $\Re S^0=0$ is needed.
Working at the $O(\epsilon)$ level and using (\ref{five}) this involves
now
\be
|\psi|^2=e^{2\Re S^1}=exp\left[\Re\sum\left(\frac{F^0_{mn}}{mn}\right)
\lambda^{-m-n}\right]
\label{OF}
\ee
\be
\Re S^0=\Re\sum_1^{\infty}T_n\lambda^n+\Re\sum_1^{\infty}S_{j+1}\lambda^
{-j}=0
\label{OG}
\ee
where one expects $S_{j+1}=-(\partial_jF^0/j)$ to be real.  This suggests
that it would be productive to think of KdV after all with $\lambda=ik$
imaginary, $T_{2n}=0,$ and $\partial_{2n}F^0=0$ as indicated below 
(so
$S_{2n+1}=0$ and only $\lambda^{-j}$ terms occur in (\ref{OG}) for $j$
odd). 
One establishes $F^0_{m,2n}=0$ as in \cite{ch} (cf. below)
so in (\ref{OF}) one only has terms
\be
\frac{F^0_{(2m+1)(2n+1)}}{(2n+1)(2m+1)}\cdot\lambda^{-2(m+n)-2}
\label{OH}
\ee
which would be real for $\lambda=ik$.  Thus $S^0$ and $P=S^0_X$ are
imaginary while $S^1$ and $P^1=\partial_X S^1$ are real. 
Note that one can make various calculations based on complex $\lambda$
or $k$ in order to establish formulas but meaning is only attached
to the formulas for $\lambda=ik$ with $k$ real.
In order to exhibit this context in a broader sense we digress
here to the Hamilton Jacobi (HJ) picture as in \cite{ci,ke}.
\\[3mm]\indent
Thus
consider the Hamilton Jacobi (HJ) theory of Section 2
in conjunction
with the formulas of Section 4.  As background let us
assume we are considering a Schr\"odinger equation which in fact
arises from a KP or KdV equation as indicated in Section 4.
Then one defines a prepotential ${\cal F}$ and it automatically must
have relations to a free energy as in (\ref{AKQ}) etc. The
HJ dynamics involve $T_n=nT'_n\,\,(\partial_n
=n\partial'_n)$ with
\be
\partial_nP=\partial{\cal B}_n+\partial_P{\cal B}_n\partial P;\,\,
\dot{P}_n=\frac{dP}{dT_n}=\partial{\cal B}_n;\,\,\dot{X}_n=\frac{dX}
{dT_n}=-\partial_P{\cal B}_n
\label{LK}
\ee
where ${\cal B}_n=\lambda^n_{+}=\sum_0^nb_{nj}P^j$
(cf. (\ref{YF}), (\ref{YG})) and this serves as a vehicle to put
$X$ in the picture in a manner commensurate with its role in
${\cal F}$.  Thereafter the HJ theory is not needed as such but
we must ask for a realistic such background theory if our insertion
of $X$ is to be meaningful.
We emphasize here the strong nature
of the dependence $\psi=\psi(X)$ and $X=X(\psi)$ with all other quantities
dependent on $X$ or $\psi$ in \cite{fa}
(along with $X=X(T)$ arising in the HJ theory) and this may
introduce constraints.
The action term $S$ is given a priori as $S(X,T,\lambda)$ with $\lambda$
given via (\ref{YC}) as a function of $P$ and we recall that $\lambda$ and
$\xi=\hat{S}_{\lambda}=S_{\lambda}-\sum_2^{\infty}nT_n\lambda^{n-1}$ are
action-angle variables with $d\lambda/dT_n=0$ and $d\xi/dT_n=-n\lambda^{n-1}$.
For the moment we do not use $dKP_{\epsilon}$ 
or $dKdV_{\epsilon}$ and it will become apparent
why they are needed.
Note that the $b_{nj}=b_{nj}(U)$ should
be real and the conditions under which the formulas of 
\cite{fa} are valid with $E=\pm\lambda^2$ real involve $\lambda$ either
real or pure imaginary.  
A little thought shows that a KdV situation here with $\lambda
=ik,\,\,\lambda^2=-k^2=-E$ would seem to work
and we try this here to see what a KdV situation 
(first without $dKdV_{\epsilon}$)
will involve. We 
will have then $P$ purely imaginary with $U_j$ and $P_j$ real
and note that only odd powers of $P$ or 
$k$ appear in (\ref{YR}).  Look now at (\ref{YR}), i.e.
$ik=P(1+\sum_1^{\infty}U_mP^{-2m})$, and for $P=iQ$ we see
that $(ik)^{2n+1}_{+}={\cal B}_{2n+1}$ will be purely imaginary.  Further
$\partial_P{\cal B}_{2n+1}$ will involve only even powers of $P$ and hence
will be real.  
Thus write now 
\be
{\cal B}_{2n+1}=\sum_0^nb_{nj}P^{2j+1};\,\,\partial_P{\cal B}_{2n+1}=
\sum_0^n(2j+1)b_{nj}P^{2j}
\label{LM}
\ee
and we have
\be
\frac{d}{dT_n}\Im{\cal F}=-\frac{1}{\epsilon}\dot{X}_n=\frac{1}{\epsilon}
\sum_0^n(2j+1)b_{nj}P^{2j}
\label{LN}
\ee
Then the condition $P=iQ$ leads to a compatible KdV situation
(\ref{LN}) and further
\be
\dot{P}_n=\frac{dP}{dT_n}=\partial{\cal B}_n=\sum_0^n\partial (b_{nj})
P^{2j+1}
\label{LO}
\ee
which is realistic (and imaginary).
\\[3mm]\indent 
Now we note that there
is danger here of a situation where $\Re P=0$ implies $\Re S=0$
which in turn would imply $|\psi|^2=1$ (going against the philosophy
of keeping $|\psi|^2$ as a fundamental variable) and this is one reason
we will need $dKdV_{\epsilon}$ with (\ref{OF}) - (\ref{OH}).
Thus in general
\be
S=\sum_1^{\infty}T_n\lambda^n-\sum_1^{\infty}\frac{\partial_n F}{n}\lambda^{-n};
\label{LP}
\ee
$$P=\lambda-\sum_1^{\infty}\frac{F_{1n}}{n}\lambda^{-n};\,\,{\cal B}_m
=\lambda^n-\sum_1^{\infty}\frac{F_{mn}}{n}\lambda^{-n}$$
and for KdV (with $\lambda=ik$) it follows from the residue formula
(cf. \cite{ch}) that
\be
F_{nm}=F_{mn}=Res_P\left(\lambda^md\lambda^n_{+}\right)
\label{LQ}
\ee
that $F_{m,2n}=0$ and from a $\bar{\partial}$ analysis (cf. \cite{ch,ci})
\be
\partial_jF=\frac{j}{2i\pi}\int\int \zeta^{j-1}\bar{\partial}_{\zeta}S
d\zeta\wedge d\bar{\zeta}
\label{LR}
\ee
The $\partial_jF$ and $F_{1j}$ can be computed explicitly as in \cite{ch}
and in particular $F_{1,2n}=0$ with ($P^2-U=-k^2$)
\be
F_{1,2n-1}=(-1)^n\left(\frac{U}{2}\right)^n\prod_1^n\frac{2j-1}{j}
\label{LS}
\ee
A further calculation along the same lines also shows that $F_{2n}=\partial_
{2n}F=0$ for KdV.
Generally $F$ will be real along with the $F_{mn}$ and we recall that the
expression for ${\cal B}_{2m+1}$ arising from (\ref{LP}) is an alternate
way of writing (\ref{LM}).  For $\lambda=ik,\,\,P$ and ${\cal B}_{2m+1}$
will be purely imaginary but $S$ could be complex 
via $\sum_1^{\infty}T_n\lambda^n$ since all powers $\lambda^n
=(ik)^n$ will occur in (\ref{LP}).  Thus $\Re S\not=0$ and we have a
perfectly respectable situation, provided the $T_{2n}$ are real.
However $T_{2n}$ imaginary as in KP1, or 
as in (\ref{BB}), would imply 
$\Re S=0$ and $|\psi|^2=1$ which is not desirable.
Another problem is that if $\Re S\not= 0$ is achieved via the times then
$|\psi|^2\sim exp[(1/\epsilon)\sum T_{2n}\lambda^{2n}]$ will not necessarily
be $\leq 1$.  Thus if $dKdV_{\epsilon}$ is not used
this would seem to force
a KP situation with $\partial_P{\cal B}_n$ real, and $P$ complex
(with constraint $\lambda=P+\sum U_{n+1}P^{-n}$ satisfying $\lambda^2$
real); then the equation $\dot{P}_n=\partial{\cal B}_n$ does not require
$\partial{\cal B}_n$ to be real.  However some care with $\lambda$ is indicated
since $P^2-U=\lambda^2\sim-k^2$ would require also $P^2$ to be real if
in fact this equation were used to define $S$ via $S_X=P$ and would force
us back to dKdV with $\Re S=0$ and $|\psi|^2=1$.  Indeed $\partial_2\psi
=\lambda^2\psi$ means $L^2_{+}\psi=\lambda^2\psi$ but we know $L^2\psi
=\lambda^2\psi$ in general so this implies $L^2=L^2_{+}$ which is KdV.
Hence we would have to go back to (\ref{A}) 
with KPI and be sure to interpret it as an
eigenvalue equation $i\hbar\partial_{\tau}\psi={\cal H}\psi =E\psi$
(we should also label $\psi=\psi_E$ as in \cite{fa},
with variable $\lambda$ divorced from $E$ entirely).
Thus one could temporarily reject dKdV, substitute dKPI, and 
continue with (\ref{LK}) with Hamiltonian type equations
$\dot{P}_n=\partial{\cal B}_n$ and $\dot{X}_n=-\partial_P{\cal B}_n$
for a Hamiltonian $H_n=-{\cal B}_n\,\,(n\geq 2)$. 
Some further argument as in \cite{cb} then shows however that dKP1 requires
$P$ genuinely complex (neither real nor imaginary) with constraints
forcing a $dKP_{\epsilon}$ theory in any event (and hence a corresponding
$HJ_{\epsilon}$ theory).  It seems that such a program would be formally
possible (modulo obstructions of a constraint nature) but the
formulas of $dKdV_{\epsilon}$ are so much simpler and 
clearly adapted to the context of \cite{fa} that it is
compelling to use them here.
\\[3mm]\indent
Thus we return to $dKdV_{\epsilon}$
and look first at the $O(\epsilon)$ version before going to a full
expansion; this seems to make the whole procedure more visible and
allows us the luxury of maintaining $\epsilon$ or $\hbar$ as a scale
parameter at the first stage.
In view of (\ref{OF}) - $(\ref{OG})$, etc., 
there is no problem with $\Re S^0=0$
while happily 
$\Re S^1\not= 0$ and $|\psi|^2\leq 1$ is realistic.  The equation
(\ref{YR}) applies now but we cannot write
$ik\sim\tilde{P}(1+q\tilde{P}^{-2})^{\frac{1}{2}}$ for $\tilde{P}=
P+\epsilon P^1$.
Indeed other terms will arise involving $P_X$ for example since,
for 
$\tilde{S}=S^0+\epsilon S^1$ with $\psi=exp(\tilde{S}/\epsilon)=exp
[(S^0/\epsilon)+S^1]$, we have $\tilde{P}=\partial\tilde{S}=S^0_X+
\epsilon S^1_X=P+\epsilon P^1$ so that $\epsilon\partial\psi=\tilde{P}
\psi=(P+\epsilon P^1)\psi,\,\,\epsilon^2\partial^2\psi=(\epsilon P_X
+\epsilon^2P_X^1)\psi +2\epsilon P^1P\psi+P^2\psi+\epsilon^2 (P^1)^2\psi$,
etc. 
along with $\epsilon\partial(\psi/\tilde{P})=
-\epsilon(\tilde{P}_X/\tilde{P}^2)\psi+\psi=\psi
-\epsilon((P_X/P^2)\psi+O(\epsilon^2)$
from which $(\epsilon\partial)\psi\to\psi$ or $(\epsilon\partial)^{-1}\psi
\to \psi/\tilde{P}$ in some sense.  Continuing such calculations we obtain
terms of $O(\epsilon)$ in $(1/\psi)(\epsilon\partial)^{-n}\psi$ of the form
$(1/\psi)
{n+1 \choose 2}(\epsilon\partial)^{-n}\left(P_X\psi/P^2\right)$
and from $L\psi=\lambda\psi$ we get to first order
$\lambda=P+\sum_1^{\infty}U_{n+1}P^{-n}+O(\epsilon)$ with a 
complicated $O(\epsilon)$ term
(cf. \cite{cb} and Section 6 for some clarification of this).
Note here also that $P
=ik-\sum_1^{\infty}P_n(ik)^{-n}$ inverts (\ref{YR}) with $P_n=0$ for
$n$ even ($P_n=F_{1n}/n$ here - cf. \cite{ch} where there is an index
shift in the $P_n$); this shows that $P=iQ$.  
Further the constraint $|\psi|^2\Im\tilde{P}=-1\equiv |\psi|^2\Im P
=-1$ and this can be written $exp(2S^1)\Im P=-1$. 
In any event this leads to expressions for $ik,\,\,(ik)_{+}^{2n+1}$, etc.
and in particular 
for $P=iQ$ imaginary and
$S^1,\,P^1$ real we obtain
$2S^1+log(\Im P)=i\pi\Rightarrow 2P^1=-(\Im P_X/\Im P)\Rightarrow
P_X=-2PP^1$.
We do not pursue this approach here however since
in fact the HJ theory is not crucial
and the methods of Section 6 will suffice.  Given $S=S^0
+\epsilon S^1$ and $F=F^0$ we know $\tilde{P}=P+\epsilon P^1$
is correct and that is all that is needed for the formulas of \cite{fa}
at the $\epsilon$ level chosen (some scaling factors still remain).
Further calculations suggest that one can obtain exact balances 
for the HJ theory (perhaps
with constraints) but higher powers of $\epsilon$ should be included
(cf. also \cite{mb,mc}); in fact the development in Section 6 should suffice
for this also, but we do not pursue the matter here.
\\[3mm]\indent
Before embarking on the full $\epsilon$ expansion let us make a few
comments on the $dKdV_{\epsilon}$ results at order $\epsilon$.
Thus we take $\lambda^2=-E$ and specify $dKdV_{\epsilon}$.  We can still
label $\psi$ as $\psi_E$ but now one can imagine a $T_2\sim\tau$ variable
inserted e.g. via $\psi=\psi(X,T_{2n+1})exp(E\tau/i\hbar)\,\,
(n\geq 0)$ with $i\hbar\psi_{\tau}=E\psi$ and $\epsilon^2\psi''-V\psi
=-E\psi=\lambda^2\psi$ where $V=V(X,T_{2n+1})$ etc.
Consider ${\cal F}=(1/2)\psi\bar{\psi}+(X/i\epsilon)$ with $\psi=
exp[(1/\epsilon)S^0+S^1],\,\,\Re S^0=0$ as in $(\ref{OG})$,
and $|\psi|^2=exp(2\Re S^1)$ as in (\ref{OF}).  Here
\be
S^0=i\left[\sum_1^{\infty}T_{2n+1}(-1)^nk^{2n+1}+
\sum_1^{\infty}S_{2n}(-1)^nk^{-2n+1}\right]
\label{OK}
\ee
and explicitly
\be
{\cal F}=\frac{1}{2}exp\left[\sum_1^{\infty}\left(\frac{F^0_{(2m+1)(2n+1)}}
{(2m+1)(2n+1)}\right)(-1)^{m+n+1}k^{-2(m+n+1)}\right]+\frac{X}{i\epsilon}
\label{OL}
\ee
Thus the $\epsilon$ ``problem" has been removed from the $|\psi|^2$ term 
in ${\cal F}$ but
$\epsilon$ still occurs as a scale factor with $X$.
Look now at (\ref{ACQ}) with $P$ replaced by $\tilde{P}$ to obtain
$|\psi|^2\Im\tilde{P}=-1$ which in view of the $\epsilon$ independence
of $|\psi|^2$ suggests that $\Im P^1=0$ which in fact is true from 
$(\ref{OH})$.
Thus $|\psi|^2\Im P=-1$ as before but $P=S^0_X$ now.  Next for $\phi=
\bar{\psi}/2\psi$ we have $\phi=(1/2)exp[-(2i/\epsilon)\Im S]$ and 
$S^0$ is imaginary as in (\ref{OK}) with $S^1$ real as indicated in 
(\ref{OH}).
Consequently
\be
\phi=\frac{1}{2}exp\left[-\frac{2i}{\epsilon}\Im S^0\right]=
\label{OM}
\ee
$$=\frac{1}{2}exp\left[-\frac{2i}{\epsilon}\left(\sum_0^{\infty} 
(-1)^nT_{2n+1}k^{2n+1}+\sum_1^{\infty}(-1)^nS_{2n}k^{-2n+1}\right)\right]$$
In the same $\epsilon$ order spirit,
one can say that $X=-\epsilon\Im {\cal F}$ and (for $P=iQ$)
\be
P=i\Im\tilde{P}=iQ=-\frac{i}{|\psi|^2}=-\frac{i}{2\Re
{\cal F}}
\label{ON}
\ee
are fundamental variables.  Note also from (\ref{OK}), $log(2\phi)=-(2/\epsilon)
S^0$, so
\be
log(2\phi)=-4i\xi=-\frac{2i}{\epsilon}\left(\sum (-1)^nT_{2n+1}k^{2n+1}+
\sum (-1)^nS_{2n}k^{-2n+1}\right)
\label{OOO}
\ee
This leads to a result from \cite{cb}, namely
\\[3mm]\indent {\bf THEOREM 5.1.}$\,\,$
From the canonical object 
$dX\wedge dP$ of Hamiltonian theory
there is a possibly fundamental symplectic form
based on the $\epsilon$ order theory, namely
\be
\frac{dX}{\epsilon}\wedge dP=
d(\Im{\cal F})\wedge\frac{i}{2(\Re{\cal F})^2}
d(\Re {\cal F})=-\frac{i}{2(\Re{\cal F})^2}d(\Im{\cal F})\wedge
d(\Re{\cal F})
\label{OP}
\ee
which 
seems intrinsically
related to the duality idea based on ${\cal F}$.
Note that this is not $dX\wedge d\tilde{P}$ (which would involve an additional
term $dX\wedge dP^1$, where a relation to $dX\wedge dP$ could then be 
envisioned via $P^1=-(1/2)\partial_Xlog\,P$).  Actually (\ref{OP})
is based only on first order WKB structure and is not dependent on KdV
connections (no ``time" dynamics is involved a priori - see also
Section 6).
\\[3mm]\indent  
The constraint $|\psi|^2\Im P=-1$ becomes
$exp[2\Re S^1]\Im S^0_X=-1$ which can be written out in
terms of $F^0=F$ and $\partial_XS_{2n}$ (cf. \cite{ca}).
Let us also compute the form 
$\omega =\delta\xi\wedge\delta\chi$ from (\ref{XX}) in
one of its many forms.  First recall $S^0$ is imaginary and $S^1$ is real
with $log(2\phi)=-(2/\epsilon)S^0=-4i\xi$ and $\chi=|\psi|^2=exp(2S^1)$.
Therefore formally, via $\xi=-(i/2\epsilon)S^0$, we have
$\omega=\delta\xi\wedge\delta\chi= -\left(i\chi/\epsilon\right)
\delta S^0\wedge\delta S^1$.
The difference here from (\ref{OP}) for example is that the term $X=-\epsilon
{\cal F}$ has no relation to $S^0$ or $S^1$ a priori.

\section{EXPANSION AND CONNECTIONS BETWEEN DUAL VARIABLES}
\renewcommand{\theequation}{6.\arabic{equation}}\setcounter{equation}{0}

Let us organize what we have so far.
From Section 3 we take a finite zone KdV
situation and produce a prepotential $F$ as in (\ref{J}) with asymptotic
connections to a BA function $\psi=exp[(1/\epsilon)S+O(1)]$ as in 
(\ref{N}) (where $\tau=exp[(1/\epsilon^2)F+O(1/\epsilon)]$ is also spelled
out).  Further one can make connections via the asymptotics of $\psi$ 
between $\Omega_n$ and ${\cal B}_n$ via $F_{mn}=q_{mn}$.  This brings
the $a_i$ variables into $F$ (and $d{\cal S}$) with ($F\sim F^0$)
\be
\partial_nF=F_n=-Res\,z^{-n}d{\cal S}=Res\,\lambda^ndS
\label{EA}
\ee
Note from $S=\sum_1^{\infty}T_n\lambda^n-\sum_1^{\infty}(\partial_jF/j)
\lambda^{-j}$ one has $dS=(\sum_1^{\infty}nT_n\lambda^{n-1}+\sum_1^{\infty}
\partial_nF\lambda^{-n-1})d\lambda$ whereas in Section 3 one is dealing
with
\be
-\partial_mF_n=F_{mn}=Res\,z^{-n}\partial_md{\cal S}=Res\,z^{-n}d\Omega_m
=-q_{mn}
\label{EB}
\ee
corresponding to $F_n=-Res\,z^{-n}d{\cal S}=F_n$.  Actually it is
interesting to compare the form of $dS$ with $d{\cal S}$ via
\be
d{\cal S}=\sum a_jd\omega_j+\sum T_nd\Omega_n=-\sum a_j(\sum\sigma_{jm}
z^{m-1})dz +
\label{ED}
\ee
$$+\sum T_n\left(n\lambda^{n-1}d\lambda-\sum q_{mn}z^{m-1}dz\right)=
\sum nT_n\lambda^{n-1}d\lambda-$$
$$-\sum z^{m-1}\left(\sum a_j\sigma_{jm}+
\sum T_nq_{mn}\right)dz$$
while $dS=\sum nT_n\lambda^{n-1}d\lambda-\sum\partial_nFz^{n-1}dz$.
Identifying $d{\cal S}$ and $dS$ we get
\be
-F_p=Res\,z^{-p}dS=Res\,z^{-p}d{\cal S}=-\sum a_j\sigma_{jp}-
\sum T_nq_{pn}
\label{EE}
\ee
which provides a formula for $F_p$ (note $\partial_na_j=0$ as indicated
in \cite{cc,ia}).  
\\[3mm]\indent
Next from Section 4 we produce ${\cal F}=(1/2)\psi\bar{\psi}+(X/i\epsilon)$
with a relation (\ref{AKQ}) between $F$ and ${\cal F}$.  Also a number
of formulas are given relating variables $\psi,\,\,\bar{\psi},\,\,
S=S^0,\,\,P=S_X=S_X^0,\,\,\phi=\bar{\psi}/2\psi=\partial{\cal F}/
\partial(\psi^2),\,\,\chi=|\psi|^2,$ and $\xi=(1/2\hbar)S$ in various
contexts.  In Section 4 the $dKdV_{\epsilon}$ theory is introduced
via $F=F^0$ in (\ref{five}), leading to $S=S^0+\epsilon S^1$ with 
$S^0$ (imaginary) in (\ref{OK}) and $|\psi|^2=exp(2\Re S^1)$ as in 
(\ref{OF}) ($S^1$ real).  The requirements of \cite{fa} produce
the constraint $|\psi|^2\Im\,P=-1$ for $P=S^0_X$ and one has fundamental
relations
\be
\phi=\frac{1}{2}exp\left(-\frac{2i}{\epsilon}\Im S^0\right)\,\,\,(cf.\,\,
(\ref{OM});\,\,X=-\epsilon\Im{\cal F};
\label{EF}
\ee
$$P=-\frac{i}{2\Re{\cal F}};\,\,log(2\phi)=-4i\xi\,\,\,(cf.\,\,(\ref{OOO})$$
plus the fundamental relation (\ref{OP}) for $dX\wedge dP$.  
In \cite{cb} a Hamilton
Jacobi theory for dispersionless theory was developed whose mission was
basically to motivate the treatment of $X$ in a canonical manner
commensurate with its role in \cite{fa}.  This is actually achieved at
the first WKB level $\psi=exp[(1/\epsilon)S+O(1)]$ but the $dKdV_{\epsilon}$
theory is needed e.g. to produce a meaningful expression for $|\psi|^2$.
\\[3mm]\indent
We now make some new computations to link various quantities.  First use
$V=-2F''$ as in (\ref{AKQ}) and recall (\ref{AA}); then (\ref{AA}) becomes
\be
\epsilon^2\psi''+2F''\psi=-E\psi
\label{EG}
\ee
Writing $\psi=exp[(1/\epsilon)S+S^1]$ this yields
\be
\epsilon^2(P^1_X+(P^1)^2)+\epsilon(P_X+2PP^1)+P^2+2F''=-E
\label{EH}
\ee
Equating powers of $\epsilon$ and recalling $P=iQ$ is imaginary with 
$P^1=S^1_X$ real one obtains
\be
F''=\frac{1}{2}(Q^2-E);\,\,P_X+2PP^1=0;\,\,P^1_X+(P^1)^2=0
\label{EI}
\ee
These equations seem to fix both $P^1$ and $P$ and thus we certainly
want
more terms via $\psi=exp[(1/\epsilon)S+S^1+\epsilon
S^2+\cdots]$ for example.  Thus using three terms we obtain
\be
P^2+2F''=-E;\,\,P_X+2PP^1=0;\,\,P_X^1+(P^1)^2+2PP^2=0
\label{EJ}
\ee
The first two equations are the same and the third shows that $P$
is not fixed by (\ref{EJ}) but a recursion procedure is suggested
determining all $P^i\sim P_i$ from $P=iQ$. 
Since relations between the $P^i\sim P_i$ here 
must agree with relations based on (\ref{LL}) or (\ref{ZZZ}) we expect
(\ref{EJ}) (expanded with $F^2$ as in (\ref{EX}) below) to be compatible
with (\ref{FN}) below for example.
Thus we will have two possible balancing procedures with expanded
$F,\,\,S$, etc. (which should agree) based on (\ref{AA}) and 
(\ref{ZZZ}) respectively.
We can also see that balancing in (\ref{ZZZ}) will require an expanded
theory at the first order WKB level.  Indeed
writing $8F''+4E=4Q^2=-4P^2$ we
obtain from (\ref{AKQ}) the equation ($F'''=-PP'=QQ'$)
\be
\epsilon^2{\cal F}'''+4Q^2\left({\cal F}'-\frac{1}{i\epsilon}\right)
+4QQ'\left({\cal F}-\frac{X}{i\epsilon}\right)=0
\label{EK}
\ee
which relates ${\cal F},\,\,P,$ and $X$. Then the
consistency of (\ref{EK}) with (\ref{EF}) relating $X,\,\,P,\,\,\Re{\cal F},$
and $\Im{\cal F}$ must be confirmed; e.g. 
$\Im{\cal F}=-(X/\epsilon)$ and $\Re{\cal F}=
-(i/2P)$ so ${\cal F}=-(1/2Q)-(iX/\epsilon)\,\,(P=iQ)$ and
(\ref{EK}) becomes (${\cal F}'-(1/i\epsilon)=Q'/2Q^2$ and ${\cal F}-
(X/i\epsilon)=-1/2Q$)
\be
\epsilon^2\left[\frac{Q'''Q-6Q''Q'+6(Q')^2}{2Q^3}\right]+4Q^2\frac{Q'}{2Q^2}
+4QQ'\left(\frac{-1}{2Q}\right)=0\Rightarrow
\label{EL}
\ee
$$\Rightarrow Q'''Q-6Q''Q'+6(Q')^3=0\equiv \left(\frac{1}{Q}\right)'''=0
\sim\frac{1}{Q}=AX^2+BX+C$$
\indent
{\bf REMARK 6.1.}$\,\,$  This is clearly absurd but it indicates that
truncation of the $\epsilon$ series will impose constraints on $Q$.
It us perhaps no surprise that a KdV connection might
restrict the WKB term $Q$ but we will see below that in fact there is no
such restriction on $Q$ in a fully expanded theory. 
We emphasize in passing that
the $(X,\psi)$ duality will be
generally meaningful for $\psi=exp[(1/\epsilon)S^0+S^1]$ with $S^0$
imaginary and $\Re S^1\not= 0$.
\\[3mm]\indent {\bf REMARK 6.2.}$\,\,$
Before developing the expanded theory let us not that
relations between the $a_i$ and ${\cal F}$
can be expected.  In the first order theory, given $F\sim F(a,T)$ as in 
(\ref{J}) one has a connection of the $a_i$ to ${\cal F}$ through $P$
via (\ref{EI}) or (\ref{EJ}) for exmple.  We recall also that $\partial_n
a_j=0$ and in the background there are Whitham equations of the form
\be
\frac{\partial d\omega_j}{\partial a_i}=\frac{\partial d\omega_i}{\partial 
a_j};\,\,\partial_nd\omega_j=\frac{\partial d\Omega_n}{\partial a_j};\,\,
\partial_nd\Omega_m=\partial_md\Omega_n
\label{EM}
\ee
(cf. \cite{cc,na}).  Given now that $d\omega_j=-\sum_1^{\infty}\sigma_{jm}
z^{m-1}dz$ along with the standard
$d\Omega_n=[-nz^{-n-1}-\sum_1^{\infty}q_{mn}z^{m-1}]dz$ the
equations (\ref{EM}) imply e.g. (cf. \cite{cn})
\be
\partial_pq_{mn}=\partial_nq_{mp};\,\,\partial_n\sigma_{jm}=\frac{\partial
q_{mn}}{\partial a_j};\,\,\frac{\partial \sigma_{jm}}{\partial
a_i}=\frac{\partial 
\sigma_{im}}{\partial a_j}
\label{EN}
\ee
In particular this indicates that $\partial_Xq_{mn}$ and $\partial_X\sigma_
{jm}$ make sense.  We could now compute $F''=F_{XX}$ from (\ref{J}) but it is
simpler to use (\ref{EE}) where
\be
F'=\sum_1^ga_j\sigma_{j1}+\sum_1^{\infty}T_nq_{1n}
\label{EQ}
\ee
from which
\be
F''=\sum_1^ga_j\sigma'_{j1}+q_{11}+\sum_1^{\infty}T_nq'_{1n}
\label{ER}
\ee
(recall from \cite{ia} that $\partial_na_j=0$).  Using (\ref{EN})
and the identification $F_{ij}=q_{ij}$ one sees that (\ref{ER}) implies
$(\clubsuit\clubsuit\clubsuit)\,\,\sum_1^ga_j(\partial F_{11}/\partial a_j)
+\sum_1^{\infty}T_n\partial_nF_{11}=0$ so $F_{11}$ is in fact homogeneous
of degree zero in $(a,T)$ (cf. \cite{cn}).  
In any event generally $F_{11}$ depends on the $a_j$
and hence so does $P$ from $P^2+2F_{11}=-E$.  Setting $P=iQ$ with
$2F_{11}=Q^2-E$ and ${\cal F}=-(1/2Q)-i(X/\epsilon)$ we see that
${\cal F}$ depends on $(a,T)$ and one can state
(as indicated in Theorem 6.8 below the result is also valid in the extended
theory for ${\cal F}\sim {\cal F}^0$)
\\[3mm]\indent
{\bf THEOREM 6.3.}$\,\,$  In the first order theory
the prepotentials ${\cal F}$ and $F$ are related
to the $a_i$ as indicated and since $\partial {\cal F}/\partial a_j=
\partial\Re{\cal F}/\partial a_j$ one has
\be
\frac{\partial{\cal F}}{\partial a_k}=\frac{\partial Q/\partial a_k}
{2Q^2}
\label{EU}
\ee
\indent
Now let us enlarge
the framework for
$F$ to $F=F_0+\epsilon F_1+\cdots$.  Thus if one assumes $F=F^0+\epsilon
F_1+\cdots\,\,(F_i\sim F^i$ here) 
for example then with $\psi=exp[(1/\epsilon)S+S_1
+\epsilon S_2+\cdots]$ (\ref{EG}) becomes at low order 
(writing
$\psi'=[(1/\epsilon)P+P_1+\epsilon P_2]\psi$ and
$\psi''=[(1/\epsilon)P'+P_1'+\epsilon P_2'+\{(1/\epsilon)P
+P_1+\epsilon P_2\}^2]\psi$ - we use $P_i$ or $P^i$ interchangeably)
$$
\epsilon^2\left[\frac{1}{\epsilon}P'+P_1'+\epsilon P_2'
+\frac{1}{\epsilon^2}P^2+P_1^2+\epsilon^2P_2^2+\frac{1}{\epsilon}
2PP_1+2PP_2+2\epsilon P_1P_2\right]+$$
\be
+2\left(F_0+\epsilon F_1+\epsilon^2F_2\right)''=-E
\label{EW}
\ee
leading to $P^2+2F_0''+E=0$ as before, plus (think of $P_{2i+1}$ as real
and $P_{2i}$ as imaginary)
\be
P'+2PP_1+2F_1''=0;\,\,P_1'+P_1^2+2PP_2+2F_2''=0;\,\,\cdots
\label{EX}
\ee
(so $F_{2i+1}$ is imaginary and $F_{2i}$ is real - we
will take $F_{2i+1}=0$ in order to have real potentials and 
to be able to 
use
arguments of \cite{ch}).
Thus $8F''+4E=\Upsilon=-4P^2-\epsilon(4P'+8PP_1)-\epsilon^2(P_1'+P_1^2
+2PP_2)+\cdots$ with ${\cal F}-(X/i\epsilon)=-(1/2\Im\,S')=-(1/2\Im\,
\tilde{P})$ where $\Im\,S'=2(Q+\epsilon\Im\,P_1+\epsilon^2\Im\,P_2+\cdots)
=\Im\tilde{P}$ and ${\cal F}'-(1/i\epsilon)=(\Im\,\tilde{P}'/2(\Im
\,\tilde{P})^2)$.  Then (\ref{AKQ}) becomes
\be
\epsilon^2{\cal F}'''+\frac{1}{2\Im\,\tilde{P}}\left[\frac{\Im\,\tilde{P}'}
{\Im\,\tilde{P}}\Upsilon-\frac{\Upsilon'}{2}\right]=0
\label{EZ}
\ee
Here $P=iQ$ and $P^2=-Q^2$ with 
\be
\Upsilon=4Q^2+\epsilon\Upsilon_1+\epsilon^2\Upsilon_2+\cdots;\,\,
\Upsilon'=8QQ'+\epsilon\Upsilon_1'+\epsilon^2\Upsilon_2'+\cdots
\label{FA}
\ee
with $\Upsilon_{2i+1}$ imaginary and $\Upsilon_{2i}$ real,
and $\Im\,\tilde{P}=Q+\epsilon{\cal P}$ 
(which should correspond to $Q+\sum_1^{\infty}\epsilon^{2i}\hat{P}_{2i}$
with ${\cal P}=\sum_1^{\infty}\epsilon^{2i-1}\hat{P}_{2i}$ and
$P_{2i}=i\hat{P}_{2i}$)
so that
\be
\frac{\Im\,\tilde{P}'}{\Im\,\tilde{P}}=\frac{Q'+\epsilon{\cal P}'}
{Q+\epsilon{\cal P}}=
\frac{Q'+\epsilon{\cal P}'}{Q}\left[1+\epsilon\frac{{\cal P}}{Q}
+\cdots\right]
\label{FB}
\ee
Hence the bracket $[\,\,\,\,]$ in (\ref{EZ}) has the form $(Q'/Q)
[4Q^2+\epsilon(\,\,\,)]-(1/2)[8QQ'+\epsilon (\,\,\,)_1]=
\epsilon
\{\,\,\,\,\}$ which leads to
\be
\epsilon^2{\cal F}'''+\frac{\epsilon}{2Q}\{\,\,\,\}=0;\,\,
{\cal F}'''=\frac{1}{2}\left[\frac{\Im\,\tilde{P}'}{(\Im\,\tilde{P})^2}
\right]''
\label{FC}
\ee
and the leading term from ${\cal F}'''$ will be the same as in (\ref{EL}).
Now the first terms
in (\ref{FC}) will involve (note ${\cal P}=\epsilon\hat{P}_2+\cdots$)
\be
{\cal F}'''=\left[\frac{1}{2}\frac{Q'+\epsilon{\cal P}'}{Q^2}\left(
1+\epsilon\frac{{\cal P}}{Q}+\cdots\right)^2\right]''=\frac{1}{2}\left[
\frac{Q'}{Q^2}+\epsilon\left(\frac{{\cal P}'}{Q^2}+\frac{Q'{\cal P}}{Q}
\right)+O(\epsilon^2)\right]''=
\label{FD}
\ee
$$=\frac{1}{2}\left(\frac{Q'}{Q^2}\right)''+\epsilon [\,\,\,]''+O(\epsilon^2)
=\frac{Q'''Q-6Q''Q'+6(Q')^3}{2Q^3}+\epsilon[\,\,\,]''+O(\epsilon^2)$$
and the first balance involves the $\epsilon$ term in $(\epsilon/2Q)
\{\,\,\,\}$ of (\ref{FC}) which can be extracted from (see below
for an expansion and note ${\cal P}=\epsilon\hat{P}_2+\cdots$)
\be
\frac{1}{2(Q+\epsilon{\cal P})}\left\{\left[\frac{Q'}{Q}+\epsilon
\left(\frac{Q'{\cal P}}{Q^2}+\frac{{\cal P}'}{Q}\right)\right]\cdot
(4Q^2+\epsilon\Upsilon_1)-\frac{1}{2}(8QQ'+\epsilon\Upsilon_1')\right\}=
\label{FE}
\ee
$$=\frac{1}{2(Q+\epsilon{\cal P})}\left\{\epsilon\frac{Q'}{Q}\Upsilon_1
-\epsilon\frac{\Upsilon'_1}{2}\right\}$$
But $\Upsilon_1=-4(P'+2PP_1)=0$ from (\ref{EX}) with $F_1=0$.  Hence
the $\epsilon$ term is automatically zero and there is no restriction
imposed here on $Q$.
We check now the next balance (which is at the same level
as (\ref{EL})).  Thus the $\epsilon^2$ term in (\ref{EZ}) will have
an $\epsilon^2$ term from (\ref{FE}) which should involve
$$
\Theta=\frac{1}{2(Q+\epsilon{\cal P})}\left\{\left[\frac{Q'}{Q}+
\epsilon\left(\frac{Q'{\cal P}}{Q^2}+\frac{{\cal P}'}{Q}\right)+
\epsilon^2\left(\frac{{\cal P}{\cal P}'}{Q^2}+\frac{Q'{\cal P}^2}{Q^2}
\right)\right](4Q^2+\epsilon\Upsilon_1+\epsilon^2\Upsilon_2)-
\right.$$
\be
\left.
-\frac{1}{2}(8QQ'+\epsilon\Upsilon_1'+\epsilon^2\Upsilon'_2)\right\}
\label{FI}
\ee
Setting ${\cal P}=\epsilon P_2+\epsilon^3P_4+\cdots$
and recalling
\be
\Upsilon_1=-4iQ'-8iQP_1;\,\,\Upsilon_2=-P_1'-P_1^2+2Q\hat{P}_2
\label{FJ}
\ee
we obtain
\be
\Theta\sim\frac{1}{2Q}\left(1+\epsilon^2\frac{\hat{P}_2}{Q}\right)\cdot\left\{
\left[\frac{Q'}{Q}+\epsilon^2\left(\frac{Q'\hat{P}_2}{Q^2}+
\frac{\hat{P}_2'}{Q}\right)
\right.\right.+
\label{FK}
\ee
$$\left.+\epsilon^4\left(\frac{\hat{P}_2\hat{P}_2'}{Q^2}+
\frac{Q'\hat{P}_2^2}{Q^2}\right)\right]
\left[4Q^2-\epsilon(4iQ'+8iQP_1)-\epsilon^2(P'_1+P_1^2-2Q\hat{P}_2)\right]-$$
$$\left.
-\frac{1}{2}\{8QQ'-\epsilon[4iQ''+8i(Q'P_1+QP_1')]-\epsilon^2[P_1''
+2P_1P_1'-2(Q'\hat{P}_2+Q\hat{P}_2')]\}\right\}$$
The $\epsilon^2$ term from $\Theta$ is then
\be
\frac{1}{2Q}\left\{
-\frac{Q'}{Q}(P_1'+P_1^2-2Q\hat{P}_2)\right.-
\label{FL}
\ee
$$-\left.4Q^2
\left(\frac{Q'\hat{P}_2}{Q^2}+\frac{\hat{P}_2'}{Q}\right)+\frac{1}{2}
[P_1''+2P_1P_1'-2(Q'\hat{P}_2+Q\hat{P}_2')]\right\}$$
so adding this to $\epsilon^2{\cal F}'''$ we require
\be
0=\frac{Q'''Q-6Q''Q'+6(Q')^3}{2Q^3}-
\label{FM}
\ee
$$-\frac{Q'}{2Q^2}(P_1'+P_1^2-2Q\hat{P}_2)-\frac{4}{2Q}(Q'\hat{P}_2
+\hat{P}'_2Q)+$$
$$+\frac{1}{4Q}[P_1''+2P_1P_1'-2(Q'\hat{P}_2+Q\hat{P}_2')]$$
Using again $Q'+2QP_1=0$ as a determination of $P_1$ with $\Upsilon_2
=-P_1'-P_1^2+2Q\hat{P}_2=2Q\hat{P}_2-(Q'/2Q)^2+(Q''/2Q)-(1/2)(Q'/Q)^2$
this yields
\be
0=\frac{Q'''Q-6Q''Q'+6(Q')^3}{2Q^3}+\frac{Q'}{2Q^2}\left[2Q\hat{P}_2+
\frac{Q''}{2Q}-\frac{3}{4}\left(\frac{Q'}{Q}\right)^2\right]-
\label{FN}
\ee
$$-\frac{2}{Q}(Q'\hat{P}_2+\hat{P}_2'Q)+\frac{1}{4Q}\left[2Q\hat{P}_2
+\frac{Q''}{2Q}-\frac{3}{4}\left(\frac{Q'}{Q}\right)^2\right]'$$
This can be then regarded as as a determination of $\hat{P}_2$ and we have
\\[3mm]\indent {\bf THEOREM 6.4.}$\,\,$ A partially expanded treatment of 
$dKdV_{\epsilon}$ theory shows that no restriction on $Q$
is required and the development will provide 
(modulo possible ``fitting" clarified below) a recursive
procedure determining the $P_i$, with first terms
$P_1=-Q'/2Q$ and $\hat{P}_2$ determined by (\ref{FN}).
\\[3mm]
\indent {\bf REMARK 6.5.}$\,\,$ Theorem 6.4 generates the $P_i$, hence
the $S_i$, and this must agree with what comes from $F=F_0+\epsilon F_1
+\cdots$.  Now we must discuss the nature of the $F_i\sim F^i$ in more
detail.  
Given $F_0$ related to KdV as above this would seem to generate
some $F_i$ via (\ref{five}),
but then a fitting
problem may arise with the requirements of Theorem 6.4
involving possibly hopeless constraints.  Thus we must
expand also the expressions based on (\ref{five}) where $F=F_0$
and consider a full $dKdV_{\epsilon}$ theory as follows.  
\\[3mm]\indent
In order to expand Remark 6.5 we consider $F=\sum_0^{\infty}\epsilon^k
F^k$ and look at the early terms.  If we remain in the context of KP
or KdV then (\ref{five}) should be implemented with
\be
F\left(T_n-\frac{\epsilon}{n\lambda^n}\right)-F(T_n)=
\sum_{k=0}^{\infty}\epsilon^k\left[-\epsilon\sum_1^{\infty}\left(
\frac{F_n^k}{n\lambda^n}\right)+\frac{\epsilon^2}{2}\sum\frac{F_{mn}^k}
{mn}\lambda^{-m-n}+O(\epsilon^3)\right]
\label{FP}
\ee
Here one is specifying $\epsilon$ as the scale factor in $T_n=\epsilon
t_n$ etc. and it is common to the expansion of $F$ and the vertex
operator calculations.  This yields then from $log\psi=(1/\epsilon)
\sum T_n\lambda^n +(1/\epsilon)\sum_1^{\infty}S_{j+1}\lambda^{-j}\sim
(1/\epsilon^2)\Delta F+(1/\epsilon)\sum_1^{\infty} T_n\lambda^n$ with
$S\to\tilde{S}=\sum_1^{\infty}T_n\lambda^n+\sum_0^{\infty}\epsilon^k
\sum_1^{\infty}S^k_{j+1}\\
\lambda^{-j}=S^0+\sum_1^{\infty}\epsilon^k
\sum_1^{\infty}S^k_{j+1}\lambda^{-j}$ and one has
$$
\frac{1}{\epsilon}\sum_0^{\infty}\epsilon^k\left[-\sum_1^{\infty}\left(
\frac{F_n^k}{n}\right)\lambda^{-n}+\frac{\epsilon}{2}\sum\left(
\frac{F_{mn}^k}{nm}\right)\lambda^{-m-n}+O(\epsilon^2)\right]=$$
\be
=\frac{1}{\epsilon}\sum_0^{\infty}\epsilon^k\sum_1^{\infty}S^k_{j+1}\lambda^{-j}
\label{FQ}
\ee
(note here that lower indices correspond to derivatives and upper indices
are position markers except for $S_{j+1}^k$ where $j+1$ is a position
marker).
Hence in particular
\be
\sum_1^{\infty}S^0_{j+1}\lambda^{-j}=-\sum_1^{\infty}\left(\frac{F^0_n}
{n}\right)\lambda^{-n};
\label{FR}
\ee
$$\sum_1^{\infty}S^1_{j+1}\lambda^{-j}=-\sum_1^{\infty}\left(\frac
{F_n^1}{n}\right)\lambda^{-n}+\frac{1}{2}\sum_1^{\infty}\left(\frac
{F_{nm}^0}{mn}\right)\lambda^{-m-n}$$
leading to
\be
S_{j+1}^0=-\frac{F^0_j}{j}\,\,\,(as\,\,\,before);\,\,S_2^k=-F_1^k;\,\,
S_{j+1}^k=-\frac{F^k_j}{j}+\frac{1}{2}\sum_1^{j-1}
\left(\frac{F^{k-1}_{m,(j-m)}}{m(j-m)}\right)
\label{FS}
\ee
for $k\geq 1$, together with
\be
\tilde{P}=\tilde{S}_X=P+\sum_1^{\infty}\epsilon^k\sum_1^
{\infty}\partial_XS^k_{j+1}\lambda^{-j}
=P+\sum_1^{\infty}\epsilon^kS_X^k;\,\,S^k=\sum_1^{\infty}S^k_{j+1}\lambda^{-j}
\label{FT}
\ee
where $S^k_{j+1}$ is given in (\ref{FS}) and $P^k=P_k=\sum_1^{\infty}
\partial_XS^k_{j+1}\lambda^{-j}$.  Now following the patterns in Section 5
we want $S^0$ imaginary with 
\be
S^0_X=P=iQ=\lambda-\sum_1^{\infty}\left(\frac{F_{1j}}{j}\right)
\lambda^{-j}
\label{FU}
\ee
while $S^1$ and $P^1=\partial_XS^1$ should be real, etc.  
and similar expansions apply for $S^k,\,\,P^k$ (cf. \ref{YR})).
\\[3mm]\indent
The spirit of KdV now gives $\partial_{2n}F^0=0$ and $F^0_{1,2n}=0$
etc. as in Section 5 (following \cite{ch}) and there seems to be no
reason why we cannot extend this to $F_{2n}=0$ and $F_{1,2n}=0$ via
$F^k_{2n}=0$ and $F^k_{1,2n}=0$, provided $F$ is real (cf. \cite{ch}).
Then as in Section 4, $P=\lambda-\sum_1^{\infty}[F^0_{1,2n-1}/(2n-1)]
\lambda^{1-2n}=iQ$, and e.g. (cf. (\ref{six}))
\be
P^1=\partial_XS^1=\partial_X\sum_1^{\infty}S^1_{j+1}\lambda^{-j}=
\label{FW}
\ee
$$=\partial_X\left[-\lambda^{-1}F_1^1+\sum_2^{\infty}\left(-\frac
{F_j^1}{j}\lambda^{-j}
+\frac{1}{2}\sum_{j\geq 2\,\,even}\lambda^{-j}\sum_1^{j-1}\left\{
\frac{F^0_{2m-1,j-2m+1}}{(2m-1)(j-2m+1)}\right\}\right)\right]$$
(the terms $F^0_{mn}$ vanish for $m$ or $n$ even so one has only
$F^0_{2m-1,2n-1}\lambda^{-2(m+n)+2}$ terms which can be labeled
as $\lambda^{-j}F^0_{2m-1,j-2m+1}$ for $j$ even).  Now $P^1$ real along
with $F$ real would be nice and (for $\lambda=ik$) 
a realization for this could
be begun via $F_j^1=0$ or simply $F^1=0$.
This situation also came up before in a pleasant way (cf. also (\ref{EX})) so
let us stipulate $F^{2i+1}=0$ and see what happens.  In particular this
drops the $F^1$ term from (\ref{FW}) and $P^1$ is then real as desired.
Further when we do this the lowest order terms involved in 
(\ref{OK}) - (\ref{OP}) remain the same but additional terms arise.
Thus consider $P\to \tilde{P}=P+\sum_1^{\infty}\epsilon^kP_k$ with $P_{2i}$
imaginary and $P_{2i+1}$ real so in (\ref{AAQ}) - (\ref{AGQ}) one replaces 
$P$ by $\tilde{P}$ and $S$ by $\tilde{S}=S^0+\sum_1^{\infty}\epsilon^k
S^k$ where we have concentrated positive powers of $\lambda$ in $S^0$.
From (\ref{FS}) we will have only $F^{2s}$ terms now which are real
and $S^k_{j+1}$ involves $F^k_j$ and $F^{k-1}_{m,(j-m)}$ so for $k=2n$ even
we have $S^{2n}_{j+1}=-F_j^{2n}/j$ while for $k=2n+1$ odd $S_{j+1}^
{2n+1}=(1/2)\sum_1^{j-1}[F^{2n}_{m,(j-m)}/m(j-m)]$ which can be rewritten
as in (\ref{FW}).  This says
\be
S^{2n} =\sum_1^{\infty}S^{2n}_{j+1}\lambda^{-j}=-\sum_1^{\infty}\left(
\frac{F_j^{2n}}{j}\right)\lambda^{-j}=-\sum_0^{\infty}\left(\frac{F^{2n}_
{2m+1}}{2m+1}\right)\lambda^{-2m-1};
\label{FX}
\ee
$$S^{2n+1}=\sum_1^{\infty}S_{j+1}^{2n+1}\lambda^{-j}=\frac{1}{2}\sum_
{j\geq 2\,\,even}\lambda^{-j}\sum_1^{j-1}\left(\frac{F^{2n}_{2m-1,j-2m+1}}
{(2m-1)(j-2m+1)}\right)$$
so $S^{2n}$ is imaginary and $S^{2n+1}$ is real for $\lambda=ik$.
Then in (\ref{AAQ}) - (\ref{AGQ}) and (\ref{ON}) - (\ref{OOO}) we have
$|\psi|^2\Im\tilde{P}=-1$ with e.g.
\be
log(2\phi)=-\frac{2i}{\epsilon}\Im\tilde{S}=\frac{2i}{\epsilon}
\left[S^0+\sum_1^{\infty}\epsilon^{2n}S^{2n}\right]
\label{FY}
\ee
Again one has $X=-\epsilon\Im
\tilde{{\cal F}}$ so (for $P_{2n}=iQ_{2n}=i\hat{P}_{2n}$)
\be
\tilde{{\cal F}}=-\frac{1}{2\Im\tilde{P}}-\frac{iX}{\epsilon};\,\,
-\frac{1}{2\Re\tilde{{\cal F}}}=\Im\tilde{P}=
Q+\sum_1^{\infty}\epsilon^{2n}Q_{2n}
\label{FZ}
\ee
where ${\cal F}\to \tilde{{\cal F}}$ represents an expansion of ${\cal F}$.
Hence in place of the essentially first order
Theorem 5.1 one would want to consider perhaps
\\[3mm]\indent {\bf THEOREM 6.6.}$\,\,$ In the fully expanded framework
just indicated one has
\be
\frac{dX}{\epsilon}\wedge d\Im\tilde{P}=\frac{dX}{\epsilon}\wedge dQ+
dX\wedge \sum_1^{\infty}\epsilon^{2n-1}dQ_{2n}=
\left(\frac{1}{2(\Re\tilde{{\cal F}})^2}\right)
d\Im\tilde{{\cal F}}\wedge d\Re\tilde{{\cal F}}
\label{GA}
\ee
(again no ``time" dynamics is involved a priori).  
\\[3mm]\indent
We note also that the potential $V$ now has the form $V=-2\partial^2_XF$
so with $F^{2k+1}=0$ and $F^{2k}$ real we have
\be
V=-2\sum_0^{\infty}\epsilon^{2k}F^{2k}_{XX}
\label{FV}
\ee
Such a formula could arise via 
$v=v(x,t_i,\alpha_k)\to v(X/\epsilon,T_i/\epsilon,a_k/i\epsilon)=
V(X,T,a)+O(\epsilon)$ in any case (as indicated after (\ref{BB}) and here
we are simply representing the $O(1/\epsilon)$ terms in $\tau=
exp[(1/\epsilon^2)F^0+O(1/\epsilon)]$ in an explicit form.
It does indicate that in order to achieve a fit between $(X,\psi)$
duality and the extended WKB theory of $dKdV_{\epsilon}$ one
must expand the potential $V$ as in (\ref{FV}).
Now look at the expanded framework and retrace the argument (\ref{EU}) - (\ref
{FN}) to see whether our procedure is adapted to determine the $F^{2n}$
with $F^{2n+1}=0$, and what is involved.  
We can also revise this procedure as in Remark 6.9 and deal with an alternative 
balancing based on (\ref{EX}), (\ref{FX}), etc.
If we take $F^1=0$ in (\ref{EX})
then $P'+2PP_1=0$ or $Q'+2QP_1=0$ and this was useful in balancing as well
as determining $P_1$ from $Q$.  Note that $Q=Q(k)$ via $Q^2=2F^0_{XX}-E$ 
where $\lambda^2=-E=-k^2$ and an expansion (\ref{YR}) holds. 
Thereafter the next balance is indicated
in (\ref{FN}) which serves to determine $P_2$ and therefrom $S^2$ and $F^2$
via 
\be
P_2=\partial_XS^2=-\sum_1^{\infty}\left(\frac{F^2_{1,2m+1}}{2m+1}\right)
\lambda^{-2m-1}
\label{GB}
\ee
Thus the $F^2_{1,2m+1}$ are in principle determined by residues from
$P_2$ and we defer momentarily the question of complete determination
of $F^2$.  The next balance arising from (\ref{EZ}) will involve the
$\epsilon$ term from $\tilde{{\cal F}}'''$ in (\ref{FD}) and the $\epsilon^3$
term in $(\epsilon/2Q)\{\,\,\,\}$ in (\ref{FC}).  Thus the $\epsilon$
term in $\tilde{{\cal F}}'''$ 
appears to be $(1/2)[({\cal P}'/Q^2)+(Q'{\cal P}/Q)]$ but
${\cal P}=\epsilon\hat{P}_2$ and hence there is no $\epsilon$ term.  For
the $\epsilon^3$ term in $(\epsilon/2Q)\{\,\,\,\}$ we go to (\ref{FC}) and 
write (recall $\Upsilon_1=0$)
\be
\frac{\epsilon}{2Q}\{\,\,\,\}=\frac{1}{2Q}\left[1+\epsilon
\left(\frac{{\cal P}}{Q}\right)+\cdots\right]\left\{\left[\frac{Q'}{Q}+
\epsilon\left(\frac{Q'{\cal P}}{Q^2}+\frac{{\cal P}'}{Q}\right)+\right.\right.
\label{GC}
\ee
$$+\left.\left.\epsilon^2\left(\frac{Q'}{Q}\left(\frac{{\cal P}}{Q}\right)^2
+\frac{{\cal P}'{\cal P}}{Q^2}\right)\right]\cdot (4Q^2+\epsilon^2\Upsilon_2)
-\frac{1}{2}(8QQ'+\epsilon^2\Upsilon'_2)\right\}$$
and one sees that there is no $\epsilon^3$ term (recall ${\cal P}\sim
\sum_1^{\infty}\epsilon^{2i-1}\hat{P}_{2i}$).  Thus the balancing act occurs
for even powers $\epsilon^{2n}$ only and will determine the $\hat{P}_{2n}$ in
terms of $Q$.  Then using (\ref{FX}) one can find $F^{2n}_{1,2m+1}$ by 
residues, and subsequently the $F^{2n}_{1,2m-1,j-2m+1}$ by differentiation,
leading to $P_{2n+1}$.  Hence (somewhat cavalierly)
\\[3mm]\indent {\bf THEOREM 6.7.}$\,\,$   The procedure indicated is
consistent and in principle allows determination of the $P_n$ and $F^{2n}$ 
from $Q$.
\\[3mm]\indent {\bf THEOREM 6.8.}$\,\,$ In the Riemann surface context
the relation $2F_0''=Q^2-E$, with $F_0''$ 
given by (\ref{ER}) ($F_0\sim F^0$) and $F=\sum\epsilon^{2n}F^{2n}$,
describes $Q$ as a function of $T_n\,\,(n\geq 2)$
and $a_j$.  Hence
by Theorem 6.7 one knows $\tilde{P}$ as a function of $T_n$ and $a_j$.
Then since $\tilde{Q}=\Im\tilde{P}=-(1/2\Re\tilde{{\cal F}})$ 
we have in place of
(\ref{EU}) the formula
\be
\frac{\partial\tilde{{\cal F}}}{\partial a_k}=\frac{\partial\tilde{Q}/
\partial a_k}{2\tilde{Q}^2}
\label{GE}
\ee
For $\tilde{{\cal F}}_0={\cal F}$ this implies the relation (\ref{EU})
of Theorem 6.3.
\\[3mm]\indent
{\bf REMARK 6.9.}$\,\,$ The balancing via (\ref{ZZZ}) as in Theorem 6.4
leading to Theorem 6.7 can be accomplished in an alternative way, which has
some simpler aspects, by working with (\ref{EX}), (\ref{FX}), etc.
Indeed, extending the calculations (\ref{EJ}) with $F=\sum F^{2n}\epsilon^
{2n}$ one obtains
\be
-Q^2+2F^0_{XX}+E=0;\,\,2QP_1 +Q'=0;\,\,P_1'+P_1^2-2Q\hat{P}_2+2F^2_{XX}=0;
\label{GF}
\ee
$$\hat{P}_2'+2QP_3+2P_1\hat{P}_2=0;\,\,P_3'-\hat{P}_2^2-2Q\hat{P}_4
+2P_1P_3+2F^4_{XX}=0;\,\,\cdots$$
Putting in power series as in (\ref{FW}) and (\ref{FX}) one can equate
coefficients of powers of $\lambda=ik$.  For example one can consider
$F^2_{11}=Q\hat{P}_2-(1/2)(P_1'+P_1^2)$ along with the expansions
\be
i\hat{P}_2--\partial_X\sum_0^{\infty}\left(\frac{F^2_{2m+1}}{2m+1}\right)
(ik)^{-2m-1}\Rightarrow \hat{P}_2=F^2_{11}k^{-1}+\cdots;
\label{GG}
\ee
$$P_1=\frac{1}{2}\sum_1^{\infty}(-1)^pk^{-2p}\sum_1^{2p-1}
\left(\frac{F^0_{2m-1,2p-2m+1}}{(2m-1)(2p-2m+1)}\right)$$
from (\ref{FW}) - (\ref{FX}), and $Q$ given via (\ref{YR}).
We have not checked the details of calculation here.
\\[3mm]
\indent {\bf REMARK 6.10.}$\,\,$  In conclusion we can say that, given
a dKdV potential $V=-2F^0_{XX}$ arising from a finite zone KdV situation
(and leading to $Q$), one can create a $dKdV_{\epsilon}$ context in which
Theorems 6.6 - 6.8 are valid.  This says that one creates both a
$dKdV_{\epsilon}$ context and an accompanying $(X,\psi)$ duality theory
in which ${\cal F}\sim \tilde{{\cal F}}$ depends on the $a_j$ variables
arising in Seiberg-Witten (SW) theory on the Riemann surface $\Sigma_g$
based on $d{\cal S}$ and $F_{{\cal S}}\sim F^0$.  In addition one makes
explicit the identifications of ${\cal F}_{{\cal S}}\sim {\cal F}_{Whitham}
\sim {\cal F}_{SW}$ with $\hat{F}$ and $F_{dKP}$ and relates these to
${\cal F}={\cal F}_{FM}$ via (\ref{AKQ}) and also via the constructions
of $\tilde{{\cal F}}$ and $\tilde{F}$ starting from $Q$.
In particular one sees also how Riemann surface information (e.g.
the $a_k$) appear in the degenerate BA function $\psi=
exp[(1/\epsilon)S+O(1)]$.
Evidently $a_j$ is not a function of $\epsilon$ and
by Theorem 6.7 the $F^{2n}$ become functions of $a_i$.  If we define
$a_i^D$ as $\partial F^0/\partial a_i$ then $a_i^D$ has no $\epsilon$
dependence but if one uses 
$a_i^D=\partial F/\partial a_i$ then $a_i^D$ acquires
an $\epsilon$ dependence
(see also e.g. \cite{cc,cg,ia} for formulas involving $a_i^D$).
In the absence of a finite zone connection
one still has all formulas indicated except for those involving the
$a_j$.  Possible ``direct" connections to quantum mechanics can arise
as indicated in Remarks 4.1 and 4.2.
Let us mention also that in terms of direct connections to SW theory
one can think of the Toda curve reformulated with a branch point at
$\infty$.  Thus e.g. $\prod_1^{2g+1}(\lambda-\lambda_i)\to
\prod_1^{2g+2}(\lambda-\lambda_j)$ and as an illustration consider
the basic elliptic curve for SW theory with $SU(2)$ (cf. \cite{cg,da,ia}).
If the corresponding one zone KdV potential $v$ 
tends to $V$ suitably then the
$(X,\psi)$ duality will be entwined with SW theory.
Here one can compare $F_{11}\sim q_{11}$ with known expressions
for $v$ in e.g. one zone KdV 
of the form $v=\sum\lambda_j-2\mu(x,t)$ and average
to obtain compatible situations (see \cite{cy} for development of this
theme).
Another intriguing possibllity here would be to develop a ``duality"
theory involving $\psi$ and $\psi^*$ for say KP; i.e. try to work with
$\psi^*=\partial {\cal F}/\partial \psi$ for some ${\cal F}$ where
e.g. ${\cal F}=(1/2)\psi\psi^*+{\cal G}$ and $\partial{\cal F}
=\psi'\psi^*$.
\\[3mm]\indent {\bf REMARK 6.11.}$\,\,$
Let us mention also the fascinating series of papers \cite{oa} by
L. Olavo, which develop quantum mechanics via the density matrix and
classical mechanics.  This was sketched briefly in \cite{cb} 
and we modify this in light of article 16 in \cite{oa} to suggest a formula
\be
\hat{{\cal F}}=\frac{1}{2}Z_Q(x,\delta x,t)
+\frac{X}{i\epsilon};
\label{OA}
\ee
$$Z_Q\sim\int F(x,,t)exp\left(\frac{ip\delta x}{\hbar}\right)dp$$
to link this framework to that of \cite{fa}.  
This involves a Wigner-Moyal infinitesimal transformation with a
phase space probability density $F(x,p,t)\sim|\psi(x,t)|^2|\phi(p,t)|^2$
for position and momentum eigenfunctions $\psi$ and $\phi$ respectively.
It seems
compelling to further develop this linkage.
In a broader sense KdV is forced upon us from the quantum mechanical (QM)
situation with a Schr\"odinger (S) equation, once the $(X,\psi)$ duality 
approach yields the Gelfand-Dickey (GD) resolvant equation.  In this
sense KdV is directly connected to QM and it would be an egregious
omission to ignore it.  In fact one should not be surprised since KdV
has already miraculously appeared in many places involving string theory,
conformal field theory (CFT), two dimensional quantum gravity (QG),
etc.  It's occurance often has a geometrical or algebraic origin based
e.g. on the Virasoro algebra, coadjoint orbits, curvature ideas, etc.
and in 2-D QG the times $t_n$ or $T_n$ are interpreted as coupling
constants (this is perhaps not unlike the fixing of times in Remark 4.1
to approximate a potential ?).  KdV is simply one of the most elementary
and important equations in all of mathematics, and incidently in mathematical
physics; its recent emergence is probably due to progress in nonlinear
mathematics as much as anything else.  That it should appear to generate
background structure for QM should be regarded as not inappropriate rather
than as a curiosity.  Another aspect of all this involves the new looks
being cast at QM itself at this historical time.  The work of Olavo for
example derives QM directly from classical statistical mechanics in a
convincing manner, leading to the assertion that QM is just an ensemble
statistical theory performed upon configuration space and related to
thermodynamical equilibrium situations.  In this approach the 
Liouville-Boltzman equation related to a WKB format and Hamilton-Jacobi
ideas play a fundamental role.  The approach in the
present paper may not be too far
removed from this.  In any event whether we start from QM and approximate,
leading to a 
background $dKdV_{\epsilon}$ theory, or begin with KdV on a Riemann
surface and create $dKdV_{\epsilon}$ with its associated $(X,\psi)$
duality theory, one is still treating $|\psi|^2$ from the associated
S equation as a fundamental variable with $|\psi|^2\leq 1$ and this
is QM.
\\[3mm]\indent {\bf REMARK 6.12.}$\,\,$
Given a Riemann surface based on KdV we find from the dispersionless
theory of dKdV at $P_{\infty}$ a potential $V=-2F_{XX}$ (coming from
$v=-2\partial^2log\tau$).  No averaging is needed to establish this 
correspondence and relations to averaging of $v$ and its derivatives are
given in \cite{bb,cc,cg,ka} for example.  This $V$ then appears in the
$(X,\psi)$ duality theory which makes sense at ground level (no
$\epsilon$ expansion) or in the expanded development of this paper.
The Riemann surface background is incidental here and the emergence of
$a_i$ variables has nothing to do with SW theory a priori.  One could however
ask what meaning accrues if the Riemann surface also comes from a Toda
theory which is related to some $N=2$ susy gauge theory based on say
$SU(n)$.  A Riemann surface based on $\prod_1^{2g+1}(\lambda-\lambda_j)$
with $P_{\infty}\sim\infty$ a branch point could conceivably also be 
represented in a form $\prod_1^{2g+2}(\lambda-\hat{\lambda})$ with two
points at $\infty$ (e.g. $\infty$ and $\hat{\infty}$ with local coordinates
$(\lambda,z)$ and $(\hat{\lambda},\hat{z})$)) as in the 
standard Toda situation
(cf. constructions in \cite{cc,co,nb}).
Recall that one can pick any $2g+2$ points $\lambda_j\in{\bf P}^1$
and there will be a unique hyperelliptic curve $\Sigma_g$ with a two fold
map $f:\,\,\Sigma_g\to {\bf P}^1$ having branch locus $\{\lambda_j\}$.  Since
any three points $\lambda_i,\,\,\lambda_j,\,\,\lambda_k$ can be sent to
$0,1,\infty$ by an automorphism of ${\bf P}^1$ the general hyperelliptic
surface of genus $g$ can be described by $2g-1$ points (moduli parameters)
on ${\bf P}^1$ (cf. \cite{cc}).  In this situation one can take $\Sigma_g$
as the spectral curve for a SW theory with action differential (cf. 
\cite{ia,na})
\be
d{\cal S}=\sum_1^ga_jd\omega_j+\sum_0^{\infty}T_nd\Omega_n+
\sum_1^{\infty}\hat{T}_nd\hat{\Omega}_n
\label{OB}
\ee
where $d\hat{\Omega}_n,\,\,\hat{T}_n\,\,(n\geq 1)$ are based at $\hat{\infty}$
and $d\Omega_0$ is a differential of third kind with simple poles at
$\infty$ and $\hat{\infty}$ having residues $\pm 1$ respectively
(cf. \cite{na} for details).  
One could then perhaps envision the emergence
of a dKdV situation with associated Schr\"odinger equation upon either
utilizing $exp(\partial/\partial m)\sim
exp(\partial m_0)[1+\epsilon\partial_X]$ 
or working at e.g. $P_{\infty}$ with $T_1$ and $T_2$ (cf. \cite{bb,nb,tb}).
Alternatively the Toda curve could be represented in a form with one
(branch point) $P_{\infty}$ as in the original $SU(2)$ case (cf. \cite{cg,ia})
and conceivably this might lead to a suitable KdV situation in some cases.
From this one could then
establish a connection to ${X,\psi}$ duality and the prepotential
${\cal F}$ of \cite{fa} as indicated in this paper (cf. also
\cite{bb,ca,na,ta}).
Note that the KdV situation gives rise via $F_{11}\sim q_{11}$ to a 
potential based on branch point moduli for example (cf. Remark 6.10),
which are automatically homogeneous of degree zero (cf. Remark 6.2
and \cite{ia}).  Subsequently one could utilize arguments as in 
Remark 4.1 to approximate a quantum mechanical potential $V$ by fixing
the values of the $T_n\,\,(n\geq 1)$.

\end{document}